\documentclass[12pt]{article}

\usepackage{graphicx}
\usepackage{color}
\usepackage{amsfonts}
\usepackage{amsmath}
\usepackage{natbib}
\usepackage{multirow}

\begin{document}
 
\newcommand{\al}{\mbox{$\alpha$}}
\newcommand{\be}{\mbox{$\beta$}}
\newcommand{\ep}{\mbox{$\epsilon$}}
\newcommand{\gam}{\mbox{$\gamma$}}
\newcommand{\sig}{\mbox{$\sigma$}}

\DeclareRobustCommand{\FIN}{%
  \ifmmode 
  \else \leavevmode\unskip\penalty9999 \hbox{}\nobreak\hfill
  \fi
  $\bullet$ \vspace{5mm}}

\newcommand{\calA}{\mbox{${\cal A}$}}
\newcommand{\calB}{\mbox{${\cal B}$}}
\newcommand{\calC}{\mbox{${\cal C}$}}

\newcommand{\muas}{\mbox{$\mu$-a.s.}}
\newcommand{\Nat}{\mbox{$\mathbb{N}$}}
\newcommand{\Rea}{\mbox{$\mathbb{R}$}}
\newcommand{\Prob}{\mbox{$\mathbb{P}$}}

\newcommand{\nin}{\mbox{$n \in \mathbb{N}$}}
\newcommand{\suc}{\mbox{$\{X_{n}\}$}}
\newcommand{\sucP}{\mbox{$\mathbb{P}_{n}\}$}}

\newcommand{\conv}{\rightarrow}
\newcommand{\convn}{\rightarrow_{n\rightarrow \infty}}
\newcommand{\convp}{\rightarrow_{\mbox{c.p.}}}
\newcommand{\convs}{\rightarrow_{\mbox{a.s.}}}
\newcommand{\convw}{\rightarrow_w}
\newcommand{\convd}{\stackrel{\cal D}{\rightarrow}}

\newtheorem {Prop}{Proposition} [section]
 \newtheorem {Lemm}[Prop] {Lemma}
 \newtheorem {Theo}[Prop]{Theorem}
 \newtheorem {Coro}[Prop] {Corollary}
 \newtheorem {Nota}{Remark}[Prop]
 \newtheorem {Ejem}[Prop] {Example}
 \newtheorem {Defi}[Prop]{Definition}
 \newtheorem {Figu}[Prop]{Figure}

\title{\sc A contamination model for approximate stochastic order: extended version.\footnote{Research partially supported by the
Spanish Ministerio de Ciencia y Tecnolog\'{\i}a, grants  
MTM2011-28657-C02-01 and MTM2011-28657-C02-02.}}

\author{Pedro C. \'Alvarez-Esteban$^{1}$, E. del Barrio$^{1}$, J.A. Cuesta-Albertos$^{2}$\\ and C. Matr\'an$^{1}$ \\
$^{1}$\textit{Departamento de Estad\'{\i}stica e Investigaci\'on Operativa and IMUVA,}\\
\textit{Universidad de Valladolid} \\ $^{2}$ \textit{Departamento de
Matem\'{a}ticas, Estad\'{\i}stica y Computaci\'{o}n,}\\
\textit{Universidad de Cantabria}}
\maketitle

\begin{abstract}
Stochastic ordering among distributions has been considered in a variety of scenarios. 
Economic studies often involve research about the ordering of investment
strategies or social welfare. However, as noted in the literature, stochastic orderings are
often a too strong assumption which is not supported by the data even in cases in which
the researcher tends to believe that a certain variable is somehow smaller than other.
Instead of considering this rigid model of stochastic order we propose to look at a more flexible
version in which two distributions are said to satisfy an approximate stochastic order relation
if they are slightly contaminated versions of distributions which do satisfy the stochastic ordering.
The minimal level of contamination that makes this approximate model hold can be used as a measure
of the deviation of the original distributions from the exact stochastic order model.
Our approach is based on the use of trimmings of probability measures. We discuss the connection between
them and the approximate stochastic order model and provide theoretical support for its
use in data analysis. We also provide simulation results and a case study for illustration.
\end{abstract}

\section{Introduction}
Stochastic order relations between distributions have been considered in a great variety of 
scenarios. Clinical studies are usually related to degrees of disease linked to 
different behaviors that often can be ordered through suitable variables. Economic studies 
frequently involve order relations on variables measuring, e.g.,  investment strategies or 
social welfare. In any case, an stochastic order  indicates a global relation between two 
distributions that improves those based on making comparisons  through individual indices 
or features of the distributions. The  stochastic order between two distributions was introduced 
in \cite{Lehmann}. In terms of distribution functions, $F, G$, it is defined by 
\begin{equation}\label{modelo0}
F \leq_{st}G\quad\mbox{ if and only if }\quad F(x) \geq G(x) \mbox{ for all } x \in \mathbb{R}
\end{equation}
(the inequality would be strict if $F(x)< G(x)$ at 
least for any $x$).  The books by \cite{Shaked} and \cite{Muller} provide 
a rather complete overview of this topic, including a discussion of a great variety 
of other stochastic orders. However, our starting point in this paper is that, as noted in \cite{Arcones}, 
these orderings are in general too strong as an assumption in problems in which one is inclined to 
believe that a population $X$ is somehow smaller than another population $Y$. In other words, 
the stochastic order is a 0-1 relation, that either holds or not. We believe that some index of 
the level of agreement with the stochastic order model for intermediate situations can be helpful.

Let us focus, for simplicity, on the two-sample problem, where two independent samples are 
obtained from $F$ and $G$. From a methodological point of view  the statistical testing problems of 
interest in relation with the stochastic order are (up to minor variations)
\begin{enumerate}\setlength{\itemindent}{2cm}
\item[a)] $H_0$: $F = G$, versus  $H_a$: $G>_{st}F$
\item[b)] $H_0$: $G \geq_{st}F$, versus  $H_a$: $G \not \geq_{st}F$
\item[c)] $H_0$: $G \not \geq_{st}F$, versus  $H_a$: $G \geq_{st}F$
\end{enumerate} 
Problem a), usually  referred to as the one-sided test, assumes that an stochastic ordering holds and 
the focus is put on giving enough statistical evidence that the relation is strict. Sometimes  such an 
assumption is scarcely justified, or it is merely the result of the intuition of practitioners. Even if 
obvious, it is relevant to say that some caution should be adopted in such cases: unlike problems b) 
or c), for arbitrary distribution functions $F$ and $G$ both $H_0$ and  $H_a$ can be false. 

Testing for stochastic dominance is the usual way of making reference to problem b),
which is the goal of a sequence of papers beginning with McFadden \cite{McFadden} and including  
\cite{Mosler}, \cite{Anderson}, \cite{Davidson}, \cite{Schmid, Schmid2},  \cite{Barrett}, 
\cite{Linton}, \cite{Linton2}, among others. 
It has the same statistical meaning as a goodness of fit problem. We just 
look for absence of evidence against our stochastic order hypothesis as a minimal requirement 
to continue our analyses under such assumption. Some weaknesses of this approach are well known 
and lead to exploring alternative or complementary tools as we will recall below.

Problem c) appears to be the most attractive for people interested in assessing the existence of an stochastic 
order between the parent distributions, because rejecting the null would provide convincing evidence to guarantee 
that $G$ stochastically dominates $F$. Unfortunately, as often happens when testing 
hypotheses, searching for a well behaved $\alpha$-level test for this problem would be unpractical: the `no data' 
test, rejecting $H_0$ with probability $\alpha$ regardless of the data is the UMP test. This is showed in \cite{Berger} 
in the one-sample setting, but the result can be easily generalized to the two-sample setup. The  
workaround used there to overcome this problem was testing `restricted stochastic dominance', that 
is, testing the property on a fixed closed interval excluding the tails of the distribution. A similar 
approach has been considered in the two-sample setting in \cite{Davidson2013}.

Here we address the problem of assessing stochastic order between two distributions as in problems b) and c) 
from a new perspective based on contaminated (mixture) models. More than an alternative technique for testing
the null models b) and c), our goal is to provide (through very simple techniques)  additional resources to 
the available procedures for the analysis of stochastic dominance. More precisely, for $\pi\in (0,1)$, we 
consider the model
\begin{equation}\label{contaminatedmodel}
F=(1-\pi)\tilde{F}+\pi H,\quad \tilde{F}\leq_{st}G,
\end{equation}
where $\tilde{F}$ and $H$ are distribution functions, and some other suitable variations of it. If model 
(\ref{contaminatedmodel}) holds true (for small $\pi$) then the stochastic order model (\ref{modelo0}) holds
except for a small fraction of observations coming from $F$ and, in this sense, we could say that
the stochastic order model is essentially valid. Alternatively, we could consider the minimal value of 
$\pi$ such that (\ref{contaminatedmodel}) holds. This yields a measure of deviation from the original
model (\ref{modelo0}). In this paper we provide appropriate statistical tools for analysis and inference
about this model as well as for suitable versions for the two-sample case.

This approach is new in this setting although it has been already 
considered in statistical testing in contingency tables and 
multinomial parametric models (\cite{Rudas} and \cite{Liu}) or in 
the analysis of similarity between samples in a 
fully nonparametric context (\cite{Alv2011b}). In fact it is closely 
related to ideas that go back to \cite{Hodges} 
and their discussion of practical vs. statistical significance.  
We further elaborate on this idea in Section 2 below.

Our handling of contaminated models is based on dealing with the dual idea of trimmings
of a probability, as introduced in \cite{Alv2008}. In general, sets of trimmings have
nice statistical properties, see, e.g., \cite{Alv2011b}. We will show that they also
behave well in the setting of stochastic ordering.

The organization of this paper is as follows. Section 2 provides some general background
on trimmings, their connections to contaminated models and, in particular, to contaminated
stochastic order models. We also discuss links to related work on approximately valid models.
In Section 3 we give asymptotic theory for approximate inference related to contaminated
stochastic order models. Finally, Section 4 contains simulation results, a real data example 
and some final conclusions.

\section{Stochastic dominance and trimming}

Trimming procedures are one of the main tools in Robust Statistics for
their adaptability  to a variety of statistical problems. By trimming according to 
a particular pattern we downplay the influence of contaminating data in our 
inferences. The introduction of data-dependent versions of trimming, often called 
impartial trimming, allows to overcome some limitations of earlier versions of 
trimming which simply removed extreme observations at tails. Generally, impartial 
trimming is based on some optimization criterion, keeping the fraction of the sample 
(of a prescribed size) which yields the least possible deviation with respect to a 
theoretical  model (see e.g.\cite{Alv2008, Alv2011b, Cuesta2, Garcia, Maronna2, 
Rousseeuw1}).

Trimming a fraction $\pi$ of a sample or data set of size $n$ 
usually means replacing the empirical measure by a new one in which the 
data are re-weighted so that the trimmed points have now zero 
probability while the remaining points will have weight $1/n(1-\pi)$. 
Instead of simply keeping/removing data we can increase 
the weight of data in good ranges (by a factor bounded by $ 1/ {(1-\pi )}$) and 
downplay the importance of data in bad zones, not necessarily removing them.
If the random generator of the sample were $P$, the theoretical counterpart 
of the trimming procedure would be to replace the probability $P(B)=\int_B1 \ dP$ 
by the new probability
\begin{equation}\label{eq1}
\tilde P(B)=\int_Bg\ dP \mbox{  where } 0\le g \le \frac 1 {(1-\pi)} \  P\mbox{-almost surely.}
\end{equation}
We call a probability measure like $\tilde P$ in (\ref{eq1}) a {\it $\pi$-trimmed} version 
or a $\pi$-trimming of $P$. The set of $\pi$-trimmings of $P$ will be denoted by $\mathcal{R}_\pi(P)$. 
If $\pi=0$ then  $\mathcal{R}_\pi(P)=\{P\}$. If $\pi=1$ then we keep the notation $\mathcal{R}_1(P)$
for the set of probabilities which are absolutely continuous with respect to $P$.
This definition of trimming has been considered by several authors 
(see, e.g., \cite{Gorda,  Alv2008}). The flexibility 
in allowing downweighting rather than removing points results in nice properties 
of $\mathcal{R}_\pi(P)$ including, in particular, a link between 
contaminated models and sets of trimmings as we show in the next result.

\begin{Prop}\label{prop1}
Let $P_0, P$ be  probability distributions on $\mathbb{R}$ with distribution functions 
$F_0$ and $F$, respectively and $\pi \in [0,1)$. The following statements are equivalent:
\begin{itemize}

\item[a)] $P=(1-\pi)P_0+\pi Q$ for some probability $Q$.
\item[b)] $(1-\pi)P_0(B)\leq P(B)$ for every measurable $B$.
\item[c)] $P_0 \in \mathcal{R}_\pi(P)$.
\item[d)] $F(x)=(1-\pi)F_0(x)+\pi G(x)$ for every $x\in \mathbb{R},$ for some  distribution function $G$.
\end{itemize}
\end{Prop}
\textbf{Proof.}  Assume a) holds so that $P=(1-\pi)P_0+\pi Q$. Since $Q$ is a probability, then $P\geq (1-\pi)P_0$ holds.
Under  condition b)  $P_0$ is absolutely continuous with respect to $P$. Hence, by the Radon-Nikodym theorem, there 
exists a nonnegative density function, say $g:=\frac {dP_0}{dP}$, such that $P_0(B)=  \int_B g\ dP$ for every $B$. 
Consider the set $B=\{g>\frac 1 {1-\pi}\}$. If $P(B)>0$ and $B_\delta=\{g\geq\frac \delta {1-\pi}\}$ then $P(B_\delta)>0$ 
for some $\delta>1$. But then
 $$(1-\pi) P_0(B)= (1-\pi) \int_B g \ dP \geq (1-\pi)  
\frac \delta {1-\pi} P(B)>P(B),$$ 
which contradicts $P\geq (1-\pi)P_0$. Therefore, $g\leq \frac 1 {(1-\pi)}$ $P-$almost surely and c) holds.

If c) holds, let $g$ be the density of $P_0$ with respect to $P$ 
which, by (\ref{eq1}), satisfies $P_0(B)=\int_B g\ dP$ and
 $0\leq g \leq \frac 1 {(1-\pi)}.$ Then $(1-\pi)P_0(B)\leq \int_B1\ dP=P(B)$ 
and we can define the nonnegative measure $\tilde{Q}(B):=P(B)-(1-\pi)P_0(B)$. Now set $Q(B):=\tilde{Q}(B)/\pi$, 
and the decomposition a) folows.
Finally note that statements a) and d) are trivially equivalent.
\hfill $\Box$

\medskip
We note that the equivalence of a), b) and c) holds in greater generality than presented here. Since we are
interested in the connection to stochastic order we refrain from pursuing
this issue here.
We note also that the contaminated model a) is not symmetric in $P$ and $P_0$. In contrast, the consideration 
of similarity between two probabilities, as introduced in \cite{Alv2011b} is a symmetric concept. We will return 
to this later in this section.

Statement d)  in Proposition \ref{prop1} involves only distribution functions which are the relevant objects in assessment 
of stochastic order. So, for the sake of economy, we will often say `$F_0$ is a trimmed version of $F$' or write 
`$F_0\in \mathcal{R}_\pi(F)$' to mean the related fact for the associated probabilities.

Let us assume that some model distribution, $F_0$, say, is given. We might be interested in assessing
whether the random generator of a sample, $F$, satisfies some stochastic order relation with respect to
$F_0$. As noted in the Introduction, model (\ref{modelo0}) is possibly a too rigid model to be realistic
and we could, instead, consider model (\ref{contaminatedmodel}), namely,
\begin{equation}\label{contaminatedmodel2}
 F=(1-\pi)G  +\pi H,\quad  \mbox{for some } G\leq_{st} F_0.
\end{equation}
Just as in Proposition \ref{prop1}, the contaminated model (\ref{contaminatedmodel2})
can be simply formulated in terms of trimming. With this goal, we write $\mathcal{F}_{st}(F_0)$ for the
set of distribution functions that are stochastically smaller than $F_0$, that is,
$$\mathcal{F}_{st}(F_0)=\{G:\, G\leq_{st} F_0\}.$$
Then (\ref{contaminatedmodel2}) holds if and only if
\begin{equation}\label{equivcontamination}
\mathcal{R}_\pi(F)\cap \mathcal{F}_{st}(F_0)\ne \emptyset
\end{equation}
and this provides an alternative description
of the contaminated model (\ref{contaminatedmodel2}) in terms of trimmings. 

So far, our contamination model deals with distributions
in an asymmetric way. We take one of them, $F_0$, as a reference model and wonder
whether the other one is, after suitable trimming, stochastically majorated by the 
model. However, with applications to two-sample problems in mind,
we should define a notion of proximity to the stochastic dominance model
on the basis of both distributions (or of both samples) without a predetermined reference model. 
In the two-sample similarity problem considered in \cite{Alv2011b}, similarity of $P_1$ and 
$P_2$ at level $\pi$ means that there exist probabilities $C, P_1',P_2'$ such that 
$P_1=(1-\pi)C+\pi P_1', P_2=(1-\pi)C+\pi P'_2$, that is, that $P_1$ and $P_2$ are $\pi$-contaminated
versions of a common distribution. This suggests that
in the present setup of stochastic order we consider the model
\begin{equation}\label{modelocontaminadotwosample}
\left\{
\begin{matrix}
F_1 & = & (1-\pi)G_1+\pi F_1'\\
& &\\
F_2 & = & (1-\pi)G_2+\pi F_2',
\end{matrix}
\right.
\qquad \mbox{ for some } G_1, G_2 \mbox{ such that } G_1\leq_{st} G_2.
\end{equation}
This contaminated model can be described, as well, in terms of trimmings. In fact, if
$\mathcal{F}_st$ denotes the set of pairs of distribution functions $(G_1,G_2)$ such
that $G_1\leq_{st} G_2$, then it follows from Proposition \ref{prop1} that 
(\ref{modelocontaminadotwosample}) holds if and only if
\begin{equation}\label{modelocontaminadotwosamplebis}
 (\mathcal{R}_\pi(F_1)\times\mathcal{R}_\pi(F_2))\cap \mathcal{F}_{st}\ne \emptyset.
\end{equation}
It is convenient at this point to assign names and notation to the contaminated 
stochastic order models.
\begin{Defi}\label{piordered} For distribution functions $F, F_0$, 
we say that $F$ is a $\pi$-contaminated version of a stochastic minorant of $F_0$ and write
$F\leq_{st,\pi}^{(1)} F_0$ if (\ref{contaminatedmodel2}) (equivalently, if (\ref{equivcontamination})) holds.
Furthermore, for distribution functions $F_1$ and $F_2$ we say that $F_1$ is
stochastically smaller than $F_2$ at level $\pi$ and write $F_1\leq_{st,\pi} F_2$ if 
(\ref{modelocontaminadotwosample}) (equivalently, if 
(\ref{modelocontaminadotwosamplebis})) holds.
\end{Defi}

With the application to two sample problems in mind, we have kept the simpler name and notation
for that case. We note also that these models can be adapted in a straighforward way to deal 
with contaminated stochastic minorization instead of majorization. We will use these models in 
the sequel with corresponding adapted notation such as $F\geq_{st,\pi}^{(1)} F_0$ or 
$F_1\geq_{st,\pi} F_2$. We note that $F_1\geq_{st,\pi} F_2$ if and only if $F_2\leq_{st,\pi} F_1$.
A bit more caution is needed for the $\geq_{st,\pi}^{(1)}$ relation, since
$F\geq_{st,\pi}^{(1)} F_0$ and $F_0\geq_{st,\pi}^{(1)} F$ are not equivalent.

We provide now simple evidence that the formulation of contaminated 
stochastic order in terms of trimmings is particularly convenient.
While two different trimmings of a fixed probability are not necessarily
comparable in stochastic order, our next result
shows that the set of trimmings of a fixed probability has a 
minimum and a maximum for the stochastic order. 
\begin{Prop}\label{prop2}
Consider a distribution function $F$ and $\pi\in [0,1)$. Define the distribution functions
\begin{eqnarray*}
F^\pi(x) &= &\left\{
\begin{array}{ll}
0 & \mbox{if } x < F^{-1}(\pi)
\\
[2mm]
\frac 1 {1-\pi} (F(x) - \pi) & \mbox{if } x \geq F^{-1}(\pi)
\end{array}
\right.
\end{eqnarray*}
and
\begin{eqnarray*}
F_\pi(x) &=& \left\{
\begin{array}{ll}
1 & \mbox{if } x \geq F^{-1}(1-\pi)
\\
[2mm]
\frac 1 {1-\pi} F(x) & \mbox{if } x <  F^{-1}(1- \pi)
\end{array}
\right.,
\end{eqnarray*}
where $F^{-1}$ denotes the quantile function associated to $F$, namely, $F^{-1}(t)=\inf \{x | \  t \leq F(x)\}$.
Then $F^\pi,F_\pi\in \mathcal{R}_\pi (F)$ and any other $\tilde{F}\in\mathcal{R}_\pi (F)$
satisfies
$$F_\pi\leq_{st}\tilde{F}\leq_{st} F^\pi.$$
\end{Prop}
\textbf{Proof.}  Consider $F_\pi$. It is easy to see that $\frac 1 \pi(F-(1-\pi)F_\pi)$ is a distribution
function, which, by Proposition \ref{prop1}, shows that $F_\pi\in\mathcal{R}_\pi(F)$.
Any other trimming of $F$, say $\tilde F$,  can be expressed (recall (\ref{eq1}))
as $\tilde{F}(x)=\int_{-\infty}^x g(t)dF(t)$ for some density $g$  (w.r.t. $P$) satisfying  
$0\le g \le \frac 1 {(1-\pi)}$. But then $\tilde{F}(x)\leq \min(\frac 1 {(1-\pi)}F(x),1)=F_\pi(x)$ for all
$x$, that is, $F_\pi\leq_{st}\tilde{F}$. The claims about $F^{\pi}$ follow similarly. \hfill $\Box$

\medskip
An interesting consequence of Proposition \ref{prop2} is that one can check whether 
the contaminated stochastic order models hold by looking just at extremes of the relevant
sets of trimmings. This, in turn, provides very simple characterizations
of the minimal contamination level required for a contaminated stochastic order model
to hold. We give details about this facts in our next results. This minimal contamination level 
under which some stochastic order relation holds is a useful measure of the deviation
from the (pure) stochastic order model and will be used in later sections. 

\begin{Prop}\label{prop3a}
For arbitrary distribution functions, $F, F_0$, and $\pi \in [0,1)$ the following are equivalent:

\smallskip
\noindent 
(a)\, $F\leq^{(1)}_{st,\pi} F_0$ \qquad \qquad \qquad \qquad \quad
(b)\, $F_\pi\leq_{st} F_0$ \qquad \qquad \qquad \qquad \quad
(c)\, $\pi\geq \pi_0$, where
$$
\pi_0:=\sup_{x:\, F_0(x)>0} \frac{F_0(x)-F(x)}{F_0(x)}.
$$
In particular, $\pi_0$ is the minimal value of $\pi$ for which $F\leq^{(1)}_{st,\pi} F_0$.
\end{Prop}

\medskip
\noindent\textbf{Proof.} If (a) holds then there exists $G\in\mathcal{R}_\pi(F_0)$ such that 
$G\leq_{st} F_0$. But then, by 
Proposition \ref{prop2}, $F_\pi\leq_{st} G\leq_{st} F_0$ and we have (b).
Obviously, (b) implies (a) since $F_\pi\in \mathcal{R}_\pi(F)$. Now, (b) is equivalent to 
$$\frac{1}{1-\pi}F(x)\geq F_0(x)$$
for every $x\in\mathbb{R}$ (note that the inequality holds trivially for $x\geq F^{-1}(1-\pi)$). But this
is, in turn, equivalent to 
$$\pi\geq \frac{F_0(x)-F(x)}{F_0(x)}$$
for all $x\in\mathbb{R}$ for which $F_0(x)>0$. This shows that (b) and (c) are equivalent too
and completes the proof. \hfill $\Box$

\begin{Nota}\label{caracterizacion}{\em In a completely symmetric fashion we could 
check that $F\geq_{st,\pi}^{(1)}F_0$ if and only if
\begin{equation}\label{stg}
\pi\geq \pi'_0=\sup_{x:\, F_0(x)<1 } \frac {F(x)-F_0(x)}{1-F_0(x)}
\end{equation}
so that $\pi'_0$ is the minimal contamination level required for $F\geq_{st,\pi}^{(1)}F_0$ to hold.}
\end{Nota}

\medskip
We deal in the next result with the $\leq_{st,\pi}$ model. It can be proved mimicking the proof of 
Proposition \ref{prop3a}, hence, we omit details.
\begin{Prop}\label{prop3b}
For arbitrary distribution functions, $F_1, F_2$, and $\pi \in [0,1)$ the following are equivalent:

\smallskip
\noindent 
(a)\, $F_1\leq_{st,\pi} F_2$ \qquad \qquad \qquad
(b)\, $(F_1)_\pi\leq_{st} (F_2)^\pi$\qquad \qquad \qquad
(c)\, $\pi\geq \pi(F_1,F_2)$, where
$$
\pi(F_1,F_2):=\sup_{x\in\mathbb{R}} (F_2(x)-F_1(x)).
$$
In particular, $\pi(F_1,F_2)$ is the minimal value of $\pi$ for which $F_1\leq_{st,\pi} F_2$.
\end{Prop}

\medskip
As in Remark \ref{caracterizacion}, we can check that $F_1\geq_{st,\pi}F_2$ if and only if
$\pi\geq \tilde{\pi}(F_1,F_2)$ $:=\sup_{x\in\mathbb{R}} (F_1(x)-F_2(x))=\pi(F_2,F_1)$. 
We note also that, since 
\begin{eqnarray*}
\sup_{x \in \mathbb{R}} |F_1(x)-F_2(x)| &&=\max \{\sup_{x \in \mathbb{R}} \left(F_2(x)-F_1(x)\right), 
\sup_{x \in \mathbb{R}} \left(F_1(x)-F_2(x)\right)\} \\&&=\max \{\pi (F_1,F_2), \pi (F_2,F_1)\},
\end{eqnarray*}
the relations $F_1\leq_{st,\pi_1} F_2$ and  $F_1\geq_{st,\pi_2} F_2$
imply $\sup_{x \in \mathbb{R}} |F_1(x)-F_2(x)|\leq\max \{\pi_1,\pi_2\}$. Hence, 
if $\pi_1$ and $\pi_2$ are small, then $F_1$ and $F_2$ are close to each other (in Kolmogorov distance).


\medskip
Next, we provide some examples 
to illustrate the meaning of the contaminated stochastic order model. 
\begin{Ejem}\label{uniforme}{\em 
Consider $F(x)=\sqrt x, \ x\in [0,1]$ and $G(x)=x, \ x \in [0,1]$. 
Then $F\leq_{st} G$. Hence, if we take $G$ to play the role of $F_0$, then 
$\pi_0=0$ or, equivalently, $F\leq^{(1)}_{st,0} G$. On the other hand, 
$$\pi'_0=\sup_{0<x<1} \frac {\sqrt{x}-x}{1-x}=\frac 1 2,$$
which means that it is necessary to trim at least 50\% of the distribution $F$ to make it
stochastically larger than $G$, that is, if we want to see $F$ as a contaminated version of
a distribution stochastically larger than $G$, then the contamination must account, at least, 
for 50\% of the distribution. With our notation, $F\geq^{(1)}_{st,\pi} G$ if and only if $\pi\geq \frac 1 2$.

If we exchange the roles of both distributions and take $F$ to play the role of $F_0$, 
there is no trimming level $\pi<1$, for which $(1-\pi)\sqrt x \leq x$.
$x\in [0,1]$ holds. This means that 
$G \leq^{(1)}_{st,\pi} F$ is impossible for $\pi \in [0,1)$ and, consequently,
we cannot see $G$ as a contaminated version of distribution stochastically smaller than $F$.

Turning our focus now to the $\leq_{st}$ relation, 
we obviously have $F\leq_{st,0} G$. Since $\sup_{x\in [0,1]}\sqrt x -x= 1/4$ we 
that $F\geq_{st,\pi} G$ if and only if $\pi\geq\frac 1 4$, that is, 
the minimum level of trimming in both distributions to reverse the original 
stochastic order is $\frac 1 4$.
\hfill $\Box$
}
\end{Ejem}

\medskip
\begin{Ejem}\label{normal}{\em 
We take now $F(x)=\Phi(x-\mu)$ and $G(x)=\Phi(x)$, where $\Phi$ is the distribution function of 
the standard normal distribution, $N(0,1)$ and $\mu>0$ ($F$ is the distribution function of the $N(\mu,1)$ law).
Obviously, $F\geq_{st} G$.
Some calculus shows that $\sup_x\frac {\Phi(x)-\Phi(x-\mu)}{\Phi(x)}=1$. Therefore,
$F \leq^{(1)}_{st,\pi} G$ is impossible if $\pi<1$ which means that the stochastic order between normal 
distributions with equal variances cannot be reversed by trimming one of them. 

The picture is different
when we move to the $\leq_{st,\pi}$ relation.
It is easy to see that
$$\pi(F,G)=\sup_{x\in \mathbb{R}} \left(G(x)-F(x)\right) =\textstyle 
\Phi\left(\frac{\mu}{2}\right)-\Phi\left(-\frac{\mu}{2}\right)
=2\Phi\left(\frac{\mu}{2}\right)-1.$$
This means that  for a shift of 0.1 units in location, we need trimming about 0.04 on both 
distributions to reverse the stochastic order. The required trimming is  0.0987 for a shift
of 0.25 units, 
0.1915 for a shift of 0.5 units and 0.3413 if the shift is of length one.
\hfill $\Box$ }
\end{Ejem}

\begin{Nota}\label{transported} {\em It is a well known fact that stochastic order
is preserved by increasing transformations, namely, if $T:\mathbb{R} \conv \mathbb{R}$
is an increasing function and $F\circ T^{-1}$ denotes the distribution function induced by $F$ 
through $T$ then $F_1\leq_{st}F_2$ implies $F_1\circ T^{-1}\leq_{st}F_2\circ T^{-1}$. This carries over
to the $\leq_{st,\pi}$ relation. In fact, let us assume that $T$ is an increasing function. 
By Proposition 2.2 in \cite{Alv2011} 
$$\mathcal{R}_\pi(F\circ T^{-1})=\{\tilde F\circ T^{-1}, \ 
\tilde F \in \mathcal{R}_\pi(F)\}.$$
Preservation of the usual stochastic order implies that the transported probabilities 
$F^\pi \circ T^{-1}$ and $ F_\pi 
\circ T^{-1}$ are the maximal and minimal, respectively, $\pi$-trimmed versions of $F\circ T^{-1}$. In 
particular, this and Proposition \ref{prop3b} imply that if $F_1\leq_{st,\pi}F_2$ then
$F_1\circ T^{-1}\leq_{st,\pi}F_2\circ T^{-1}$. As a trivial consequence, for instance, 
if $F$ and $G$ are the distribution functions of lognormally distributed random variables $X=\exp (N 
+ \mu)$ and $Y=\exp (N)$, where $N$ is a standard normal random variable and $\mu>0$, then 
$F\geq_{st} G$ and, by Example \ref{normal}, $F\leq_{st,\pi}G$ if and only if
$\pi\geq2\Phi\left(\frac{\mu}{2}\right)-1$.\quad $\Box$ }
\end{Nota}

\medskip
We close this section with a comparison to alternative approaches to a \textit{relaxed} 
version of the stochastic order. In \cite{Arcones}, the value $P(X\leq Y)$, where $X$ and $Y$ 
are independent random variables with distribution functions 
$F$ and $G$, is considered as an index of \textit{precedence} of $F$ to $G$. The relation $F\leq_{sp}G$ ($F$ stochastically
precedes to $G$) is defined by $\theta(F,G)\geq1/2$, where $$\theta(F,G):=P(X\leq Y)=\int (1-G(x-))dF(x).$$ This is motivated
by the fact that, if $F\leq_{st}G$
then $\theta(F,G)\geq 1/2$, with strict inequality unless $F=G$.  On the other hand, this index can be greater 
than $1/2$, even if $F\not \leq_{st}G$. In other words,  $F\leq_{st}G$ is a stronger relation than $F\leq_{sp}G$. It is argued then
that the weaker nature of the relation $F\leq_{sp}G$ can be more versatile in some applications.

As an illustrative example in \cite{Arcones} the authors note  that, if $F(x)=\Phi\left(\frac {x-\mu}\sigma \right), G(x)=
\Phi\left(\frac {x-\nu}\tau \right)$ are the distribution functions of two normal laws, then $F\leq_{st}G$ if and only if 
$\mu \leq \nu$ and $\sigma=\tau$, while $F\leq_{sp}G$ as soon as $\mu \leq \nu$. However, we note that this precedence 
relation leads to consider that a distribution stochastically precedes that degenerated on its median, while, in fact, just 
half of  the times it will produce values below its own median. 

In contrast, and we think that more in line with the announced goal of giving a sound treatment to statements like
`We believe that population $X$ is somehow smaller than population $Y$', we note that the relation $\leq_{st,\pi}$
allows to  assess to what extent it can be expected that the values obtained from the first distribution will be smaller 
than those obtained from the second.

\section{Inference in contaminated stochastic order models}

In this section we will assume that  $X_1,...,X_n$ and $Y_1,...,Y_m$ are independent 
i.i.d. random samples obtained from $F$ and $G$, respectively. Our goal is to 
provide statistical methods for the assesment of contaminated stochastic order
between $F$ and $G$. More specifically, we are interested in testing the null model
\begin{equation}\label{nullmodel}
H_0:\, F\leq_{st,\pi}G,
\end{equation}
for a fixed value of $\pi$ against the alternative $H_a:\,F\not \leq_{st,\pi} G$. 
We are also interested in estimation of the minimal contamination
level under which $F\leq_{st,\pi}G$ holds. The methods to be presented in this section 
could be easily adapted for inference about the $\leq_{st,\pi}^{(1)}$ order. For the sake of
brevity we refrain from pursuing this issue in this paper.

From Proposition \ref{prop3b} it is clear that
the testing problem and the estimation problem are very 
closely related, since we can rewrite (\ref{nullmodel}) as the problem
of testing
\begin{equation}\label{nullmodelbis}
H_0:\, \pi(F,G)\leq \pi
\end{equation}
against the alternative $H_a:\, \pi(F,G)> \pi$. Therefore, 
if $\hat{L}=\hat{L}(X_1,...,X_n,Y_1,...,Y_m)$ were an (asymptotic) lower confidence
bound for $\pi(F,G)$, rejection of $H_0$ when $\hat{L}\geq \pi$ would yield a test with
(asymptotically)  controlled type I error probability. We note also that often the real goal
of the researcher will be  \textit{to conclude that stochastic order essentially holds}. But then,
within the present setup, the testing problem under consideration should be
\begin{equation}\label{nullmodel2}
H_0:\, \pi(F,G)\geq \pi
\end{equation}
against $H_a:\, \pi(F,G)< \pi$. Rejection of (\ref{nullmodel2}) would provide statistical evidence
that stochastic order, up to some (hopefully small) contamination, holds. In this case we could
base our decission on an upper confidence bound for $\pi(F,G)$.

\medskip
We will use the empirical version as the estimator of $\pi(F,G)$. More precisely, we write 
$F_n$ and $G_m$, respectively, for the sample distribution functions of the $X$'s and 
$Y$'s samples and take 
$$\pi(F_n,G_m)=\sup_{x\in\mathbb{R}}(G_m(x)-F_n(x))$$
as estimator of $\pi(F,G)$. We will make the following assumption in this section.

\bigskip
\noindent {\bf Assumption A1:}  $F$ and $G$ are continuous, and   $n,m \to \infty$ 
in such a way that $\lambda_{n,m}:=\frac n {n+m} \to 
\lambda \in (0,1)$.
\vspace{3mm}

From the Glivenko-Cantelli theorem, we trivially obtain that $\pi(F_n,G_m)$ is a consistent estimator, namely,
\begin{equation}\label{G-C}
\pi(F_n,G_m) \to \pi(F,G) \mbox{ almost surely, as } m, n \to \infty.
\end{equation}
Under the homogeneity hypothesis $F=G$, it is well-known (see e.g. \cite{Durbin})
that $\pi(F_n,G_m)=\sup_{x \in \mathbb{R}}(F_n(x)-G_m(x))$ is distribution free and, 
furthermore, that
\begin{equation} \label{Smirnov}
{\textstyle \sqrt \frac {mn}{m+n}} \pi(F_n,G_m) \rightarrow_w \bar{B}:=\sup_{t \in [0,1]}B(t),
\end{equation}
where $B(t)$ denotes a Brownian bridge on $[0,1]$. This result allows easy computation of asymptotic critical values
for testing the null model $F=G$. In fact, $P(\bar{B}>x)=\exp(-2x^2), x \geq 0$.
On the other hand, for $F\ne G$, $\pi(F_n,G_m)$ is no longer distribution free, not even asymptotically 
as we can see in the next result. This suggests that we consider a bootstrap version of $\pi(F_n,G_m)$ as follows.
Given $X_1,\ldots,X_n$ we take $X_1^*,\ldots,X_n^*$ to be i.i.d. with common ditribution $F_n$ and
write $F_n^*$ for the empirical distribution on $X_1^*,\ldots,X_n^*$. Similarly we
define $G_m^*$. With this setup we have the following.

\begin{Theo}\label{Ragha}
Under assumption A1, if we denote $\Gamma(F,G):=\{x\in\mathbb{R}: 
G(x)-F(x)=\pi(F,G) \}$ and  $B_1(t)$ and $ B_2(t)$  are independent Brownian 
Bridges on $[0,1]$, then
\begin{equation}\label{asympt}
{\textstyle \sqrt \frac {mn}{m+n}} \left(\pi(F_n,G_m)-\pi(F,G)\right) \convw \sup_{x \in 
\Gamma(F,G)}\left(\sqrt{\lambda} \ B_1(G(x))-\sqrt{1- \lambda} \  B_2(F(x))\right).
\end{equation}
Furthermore, if $\delta_{n,m}=K\sqrt{\frac{n+m}{nm}\log\log(\frac{nm}{n+m})}$, $K>2$ and 
$\Gamma_n(F_n,G_m)=\{x:\, G_m(x)-F_m(x)\geq \pi(F_n,G_m)- \delta_{n,m}\}$
then, conditionally given the $X_i$'s and $Y_{j}$'s, 
\begin{eqnarray}\nonumber 
\lefteqn{{\textstyle \sqrt \frac {mn}{m+n}} \sup_{x\in \Gamma_n(F_n,G_m)}((G_m^*(x)-G_m(x))-(F_n^*(x)-F_n(x))))}
\hspace*{4cm}\\
\label{asymptboot}
&\convw &\sup_{x \in 
\Gamma(F,G)}\left(\sqrt{\lambda} \ B_1(G(x))-\sqrt{1- \lambda} \  B_2(F(x))\right)
\end{eqnarray}
with probability one.
\end{Theo}

\begin{Nota}\label{NotaDistLimite}{\em
The convergence result in (\ref{asympt}) is just a rewriting of Theorem 4  in \cite{Ragh73}.
In the Appendix we give a proof that yields (\ref{asymptboot}) with little additional effort.
We note that (\ref{asympt}) includes (\ref{Smirnov}) because for 
equal distributions $F=G$, we have $\pi(F,G)=0$ and $\Gamma(F,G)=\mathbb{R}$. We observe
further that with the alternative notation 
$$T(F,G,\pi):= \{t \in [\pi,1]: \ G(x)=t \mbox{ and } F(x)=t-\pi  
\mbox{ for some } x \in \mathbb{R} \},$$
the limit law in (\ref{asympt}) can be rewritten
\begin{equation}\label{alternative}
\bar{B}(F,G,\lambda):=\sup_{t \in T(F,G,\pi(F,G))}\left(\sqrt{ \lambda} \ 
B_1(t)- \sqrt{1-\lambda} \ B_2(t-\pi(F,G))\right).
\end{equation}
If $T(F,G,\pi(F,G))$ consists of a single point, say $t_0$ (note that $t_0\in[\pi(F,G),1]$ in that case),  then 
$\bar{B}(F,G,\lambda)$ is centered normal with variance
$\lambda t_0(1-t_0)+(1-\lambda)(t_0-\pi(F,G))(1-t_0+\pi(F,G))$. If $T(F,G,\pi(F,G))$
contains two or more points then $\bar{B}(F,G,\lambda)$ is not normal and has positive expectation.
In fact, it is the possibility of having two or more points in $T(F,G,\pi(F,G))$ which
has motivated the bootstrap proposal in Theorem \ref{Ragha} instead of simply
considering the \textit{direct} bootstrap version
$${\textstyle \sqrt \frac {mn}{m+n}} \left(\pi(F^*_n,G^*_m)-\pi(F_n,G_m)\right).$$
A look at the proof of Theorem \ref{Ragha} in the Appendix shows that the conditional
asymptotic beahavior of
${\textstyle \sqrt \frac {mn}{m+n}} \left(\pi(F^*_n,G^*_m)-\pi(F_n,G_m)\right)$ 
mimmicks that of ${\textstyle \sqrt \frac {mn}{m+n}} (\pi(F_n,G_m)-\pi(F,$ 
$G))$ (hence the bootstrap works)
if $T(F,G,\pi(F,G))$ consists of only one point but can behave differently 
otherwise. See also Proposition ??? below.
}
\end{Nota}

It is convenient at this point to introduce the notation
\begin{equation}\label{alternativeextra}
\bar{B}(a,\lambda):=\sup_{t \in [a,1]}\left(\sqrt{ \lambda} \ 
B_1(t)- \sqrt{1-\lambda} \ B_2(t-a)\right).
\end{equation}
It is easy to see that $\bar{B}(a,\lambda)$ is a particular case of $\bar{B}
(F,G,\lambda)$, coming, for instance, from the choice $F$ (respectively $G$)
equal to the distribution function of the uniform distribution on $[a,1]$ (resp.
uniform on $[0,1]$). 
The next result provides simple but useful upper and lower bounds for the quantiles 
of $\bar{B}(F,G,\lambda)$.

\begin{Prop}\label{quantilebounds} If $\alpha\in(0,1)$, $K_\alpha(F,G,\lambda)$ (resp. $K_\alpha(a,\lambda)$)
is the $\alpha$-quantile of $\bar{B}(F,G,$ $\lambda)$ (resp. $\bar{B}(a,\lambda)$)
and $\Phi$ denotes the standard normal distribution function then
$$K_\alpha(F,G,\lambda)\leq K_\alpha(\pi(F,G),\lambda).$$
Furthermore, if $\alpha\in[\frac 1 2,1)$
$$\bar{\sigma}(F,G,\pi(F,G))\Phi^{-1}(\alpha)\leq K_\alpha(F,G,\lambda)$$
where $\bar{\sigma}(F,G,\pi(F,G))=\max_{t\in T(F,G,\pi(F,G))} \sigma_t$ and
$\sigma_t^2=\lambda t(1-t)+(1-\lambda)(t-\pi(F,G))(1-t+\pi(F,G))$, while for
$\alpha\in(0,\frac 1 2)$
$$\underline{\sigma}(F,G,\pi(F,G))\Phi^{-1}(\alpha)\leq K_\alpha(F,G,\lambda)$$
with $\underline{\sigma}(F,G,\pi(F,G))=\min_{t\in T(F,G,\pi(F,G))} \sigma_t$
\end{Prop}

\medskip
From Proposition \ref{quantilebounds} we see that quantiles of $\bar{B}(F,G,\lambda)$
are bounded below by normal quantiles.
Optimization of $\sigma_t^2$ over the interval $[\pi(F,G),1]$ shows that
$$\underline{\sigma}_{\pi(F,G)}
\leq\underline{\sigma}(F,G,\pi(F,G))\leq \bar{\sigma}(F,G,\pi(F,G))\leq \bar{\sigma}_{\pi(F,G)},$$
with $\underline{\sigma}_{\pi}^2=\min(\lambda,1-\lambda)\pi(1-\pi)$, 
\begin{equation}\label{sigmabar}
\bar{\sigma}_{\pi}^2=\left\{
\begin{matrix}
\frac{1}{4}-\pi^2\lambda(1-\lambda)& & \mbox{if }\lambda\pi\leq \frac 1 2 \mbox{ and }(1-\lambda)\pi\leq \frac 1 2,\\
\lambda \pi(1-\pi) & & \mbox{if } \lambda\pi>\frac 1 2,\\
(1-\lambda) \pi(1-\pi)& & \mbox{if } (1-\lambda)\pi>\frac 1 2.
\end{matrix}
\right.
\end{equation}
This entails
that for $\alpha\in[\frac 1 2,1)$
\begin{equation}\label{sigmamin}
\underline{\sigma}_{\pi(F,G)}\Phi^{-1}(\alpha)\leq K_\alpha(F,G,\lambda)
\end{equation}
and for $\alpha\in(0,\frac 1 2)$
\begin{equation}\label{sigmamax}
\bar{\sigma}_{\pi(F,G)}\Phi^{-1}(\alpha)\leq K_\alpha(F,G,\lambda).
\end{equation}

\medskip
On the other hand, upper bounds for quantiles of
$\bar{B}(F,G,\lambda)$ are given by  quantiles of
$\bar{B}(\pi(F,G),\lambda)$. We provide next a
simpler representation (in distribution)
$\bar{B}(a,\lambda)$, in terms of only one 
Brownian bridge and use it to derive a useful 
expression for the computation of its quantiles. We also give
a simple expression for the mean and the variance. Note 
that to avoid confusion we state the result for $\bar{B}(a,\lambda)$, with
$\pi$ denoting the usual constant in the following equations.

\begin{Prop}\label{DistRepre}
\begin{itemize}
\item[(a)] If $a\in (0,1)$ and $u\in\mathbb{R}$
\begin{eqnarray*}
\lefteqn{P(\bar{B}(a,\lambda)>\sqrt{1-a}u)=1-\Phi({\textstyle \frac{u}{\sqrt{\lambda a}}})\Phi({\textstyle \frac{u}
{\scriptstyle \sqrt{\scriptstyle (1-\lambda) a}}})+ e^{-\frac{2(1-a) u^2 }{1-4\lambda (1-\lambda)a^2}}}
\hspace*{0.5cm}
\\
&&
\times \int_{-\infty}^{\frac u {\sqrt{\lambda a}}} {\textstyle \frac 1 {\sqrt{2\pi}}} e^{-\frac{1-4\lambda (1-\lambda)a^2}{2}(x-
\frac{2u\sqrt{\lambda a}(1-2(1-\lambda)a)}{1-4\lambda (1-\lambda)a^2})^2} \Phi\left({\textstyle \frac{u(1-2(1-\lambda)a)}
{\sqrt{(1-\lambda)a}}+{\scriptstyle 2\sqrt{\lambda (1-\lambda)}} ax } \right) dx.
\end{eqnarray*}
\item[(b)] If $a\in [0,1)$ 
$$E(\bar{B}(a,\lambda))={\textstyle \frac 1 {\sqrt{2\pi}}}\left[\sqrt{a(1-a)}+{\textstyle\frac{\pi}2} -\mbox{\em atan} 
\left({\textstyle \sqrt{ \frac {a}{1-a}} }\right)\right],$$
$$\mbox{\em Var}(\bar{B}(a,\lambda))={\textstyle\frac{1-a^2}2 }+\pi |2\lambda -1|a -{\textstyle \frac 1 
{{2\pi}}}\left[\sqrt{a(1-a)}+{\textstyle\frac{\pi}2} -\mbox{\em atan} 
\left({\textstyle \sqrt{ \frac {a}{1-a}} }\right)\right]^2.$$
\end{itemize}
\end{Prop}

Part (a) of the last result easily yields that $\bar{B}(a,\lambda)$ has subgaussian tails, with, for instance,
$P(\bar{B}(a,\lambda)>t)\leq 3 e^{-\frac{ 2t^2}{1-4\lambda(1-\lambda)a^2} }$ for $t>0$. It can also be used to compute
approximately probabilities and quantiles of $\bar{B}(a,\lambda)$ through numerical integration. We return to this
issue in Section 4. From (b) we see that $E(\bar{B}(a,\lambda))$ as a function of $a$ (it does not depend on $\lambda$) 
decreases from $\sqrt{\frac{\pi}8}$, for $a=0$ to $0$ as $a\to 0$ and also that an unbalanced design 
($\lambda\ne \frac 1 2$) results in an increase in variance, more important for large values of $a$.

\subsection{Testing for essential stochastic order}
We turn here to the testing problem (\ref{nullmodel2}), namely,
\begin{equation}\label{nullmodel2b}
H_0:\, \pi(F,G)\geq \pi_0 \quad \mbox{vs.} \quad H_a:\, \pi(F,G)< \pi_0
\end{equation}
for a fixed $\pi_0\in(0,1)$. We reject $H_0$ for small values of $\pi(F_n,G_m)$.
More precisely, if $\alpha<\frac 1 2$, our first proposal is rejection of $H_0$
in (\ref{nullmodel2b}) if 
\begin{equation}\label{proposal1}
\textstyle{\sqrt{\frac{nm}{n+m}}
(\pi_{n,m}-\pi_0) < \bar{\sigma}_{\pi_0}\Phi^{-1}(\alpha)},
\end{equation}
with $\bar{\sigma}_\pi$ as in (\ref{sigmabar}). We show next that
this provides a test of asymptotic level $\alpha$, which detects alernatives
with power exponentially close to one (see Remark \ref{expunifconsistent} below). The result identifies
a least favorable pair within $H_0$.
For a cleaner statement we will
write $\pi_{m,n}$ for $\pi(F_n,G_m)$ and $\mathbb{P}_{F,G}$ for the probabilities under the
assumption that the underlying distribution functions of the two samples are $F$ and $G$, respectively.

\begin{Prop}\label{Test1}
With the above assumptions and notation, if $\alpha<\frac 1 2$ and $\pi_0\leq \frac 1 2$,
\begin{eqnarray}\nonumber
\lefteqn{\lim_{n\to\infty} \sup_{(F,G)\in H_0} \mathbb{P}_{F,G}(  \textstyle{\sqrt{\frac{nm}{n+m}}
(\pi_{n,m}-\pi_0) < \bar{\sigma}_{\pi_0}\Phi^{-1}(\alpha)} )}\hspace*{1cm}\\
\label{asymptoticlevel}
&=& \lim_{n\to\infty}  \mathbb{P}_{F_0,G_0}(  \textstyle{\sqrt{\frac{nm}{n+m}}
(\pi_{n,m}-\pi_0) < \bar{\sigma}_{\pi_0}\Phi^{-1}(\alpha)} ) =\alpha,
\end{eqnarray}
where $F_0$ is the distribution function of the law 
$(\frac 1 2-\pi_0\lambda)U(0,\frac 1 2+\pi_0(1-\lambda))+ 
(\frac 1 2 +\pi_0\lambda)U(\frac 1 2+\pi_0(1-\lambda),1+\frac {\pi_0} 2-\lambda \pi_0^2)$ 
and
$G_0$ is the distribution function of the law $U(0,1)$. Furthermore, if $\pi(F,G)>\pi_0$ then
\begin{equation}\label{errortipo1}
\mathbb{P}_{F,G}(\textstyle{\sqrt{\frac{nm}{n+m}}
(\pi_{n,m}-\pi_0) < \bar{\sigma}_{\pi_0}\Phi^{-1}(\alpha)})
\leq e^{-2\frac{nm}{n+m}(\pi_0-\pi(F,G))^2},
\end{equation}
while if $\pi(F,G)<\pi_0$ then for $n,m$ such that $\textstyle {\frac {nm}{n+m}}(\pi_0-\pi(F,G))^2\geq  (\frac{1}2 +
\sqrt{\lambda_n(1-\lambda_n)})\log 2 $ $-\bar{\sigma}_{\pi_0}\Phi^{-1}(\alpha)$,
we have
\begin{equation}\label{power}
\mathbb{P}_{F,G}(  \textstyle{\sqrt{\frac{nm}{n+m}}
(\pi_{n,m}-\pi_0) \geq \bar{\sigma}_{\pi_0}\Phi^{-1}(\alpha)} )\leq 2 e^{-\frac{2}{1+2
\sqrt{\lambda_n(1-\lambda_n)}} (\bar{\sigma}_{\pi_0}\Phi^{-1}(\alpha)
+\sqrt{\frac {nm}{n+m}}(\pi_0-\pi(F,G)))^2}.
\end{equation}
\end{Prop}

\begin{Nota}\label{expunifconsistent}{\em 
Proposition \ref{Test1} means that we can test $H_0: \pi(F,G)\geq \pi_0$ with a test
of asymptotic level $\alpha$ and furthermore, that alternatives, $\pi(F,G)<\pi_0$,
can be detected with power exponentially close to one. In fact, focusing
for simplicity in the case $m=n$, we see from (\ref{power}) that if $C<1$ then for 
large enough $n$ any alternative 
$\pi(F,G)\leq \pi_1<\pi_0$
will be rejected with power at least $1-2 e^{-\frac C 2 (\pi_0-\pi_1)^2n}$. If we combine this with 
(\ref{errortipo1}) we see that our proposal yields a test of $H_0': \pi(F,G)\geq \pi_0' $ against
$H_a': \pi(F,G)\leq \pi_1$ (with $\pi_0'>\pi_0>\pi_1$) which is uniformly exponentially consistent
in the sense that both type I and type II error probabilities are uniformly
exponentially small for large enough $n$. We refer to \cite{Barron89} for further discussion
on uniformly exponentially consistent tests.
}
\end{Nota}

\begin{Nota}{\em It is not really necessary to consider only the case $\pi_0\leq \frac 1 2$
in Proposition \ref{Test1}. In fact the same holds if 
$\pi_0\lambda \leq \frac 1 2$ and $\pi_0(1-\lambda) \leq \frac 1 2$, which is always
the case if $\pi_0\leq \frac 1 2$. Otherwise, if $\pi_0>\frac 1 2$ we could have
$\pi_0\lambda >\frac 1 2$ or $\pi_0(1-\lambda) >\frac 1 2$. In the first case
Proposition \ref{Test1} holds if we take $F_0$ to be the distribution function 
of the law $U(\pi_0,1)$ and in the second we have to take the law 
$(1-\pi_0)U(0,1)+\pi_0 U(1,1+\pi_0(1-\pi_0))$. Details can be checked in a straighforward
way. From an applied point of view the interest of Proposition \ref{Test1} is to assess that
stochastic order holds up to some small contamination, say $\pi_0=0.1$,  $\pi_0=0.05$
or $\pi_0=0.01$. For this reasons and to get a simpler statement we have chosen
to focus on the case $\pi_0\leq \frac 1 2$.
}
\end{Nota}

\medskip
While Proposition \ref{Test1} guarantees fast convergence to 0 of type II error probabilities
and of type I error probabilities as we move away from the boundary with the test in 
(\ref{proposal1}), the test is somewhat conservative for finite samples as we will see in
Section 4 and some alternative procedures can be of interest. One possibility is to reject 
$H_0:\, \pi(F,G)\geq \pi_0$ if
\begin{equation}\label{proposal2}
\textstyle{\sqrt{\frac{nm}{n+m}}
(\pi_{n,m}-\pi_0) < \hat{\sigma}_{n,m}\Phi^{-1}(\alpha)},
\end{equation}
where $\hat{\sigma}_{n,m}^2=\min_{t\in T(F_n,G_m,\pi(F_n,G_m))} \sigma_t^2$
and $\sigma_t^2$ is as in Proposition \ref{quantilebounds} replacing
$F$ with $F_n$ and $G$ with $G_m$. A little thought shows that this increases (slightly) the probability
of rejection at the boundary (if $\alpha<\frac 1 2$ we are increasing the cut value), while providing 
still a test of asymptotic level $\alpha$. 
In Section 4 we show the improvement that (\ref{proposal2}) provides over (\ref{proposal1}) for 
finite samples.

\medskip
A second, more important source of improvement comes from the consideration of bias corrected
estimators instead of $\pi(F_n,G_m)$. In fact 
$$E(\sup_x (G_m(x)-F_n(x))\geq \sup_x E(G_m(x)-F_n(x))= \sup_x(G(x)-F(x)).$$
This implies $\mbox{bias}(\pi(F_n,G_m))=E(\pi(F_n,G_m))-\pi(F,G)\geq 0$. 
Furthermore,
it is easy to see from the proofs in the Appendix that
$${\textstyle\sqrt{\frac{mn}{m+n}} }\mbox{bias}(\pi(F_n,G_m)) \to E\Big(\sup_{t\in T(F,G,\pi(F,G))}
\sqrt{\lambda} B_1(t)-\sqrt{1-\lambda}B_2(t-\pi(F,G))\Big).$$
Combining this last display with (b) in Proposition \ref{DistRepre} and subsequent comments,
we see that, asymptotically, $\mbox{bias}(\pi(F_n,G_m))$ is smaller than 
$\sqrt{\frac{\pi}8}\sqrt{\frac{m+n}{mn}}\simeq 0.63 \sqrt{\frac{m+n}{mn}}$. We also see 
that $\sqrt{\frac{m+n}{mn}}\mbox{bias}(\pi(F_n,G_m))\to 0$ when $\pi(F_n,G_m)$ is asymptotically normal
(that is, when $T(F,G,\pi(F,G))$ consists of a single point).

\medskip
We consider the bootstrap bias estimator
$$\widehat{\mbox{bias}}_{\mbox{\scriptsize BOOT}}(\pi(F_n,G_m)):=E^*(\pi(F_n^*,G_m^*))-\pi(F_n,G_m),$$
where $F_n^*$, $G_m^*$ are as in Theorem \ref{Ragha} and $E^*$ denotes conditional expectation given the
$X_i$'s and $Y_j$'s. We define the bias corrected estimator
\begin{equation}\label{biascorrected}
\hat{\pi}_{n,m,\mbox{\scriptsize BOOT}} :=\pi(F_n,G_m)-\widehat{\mbox{bias}}_{\mbox{\scriptsize BOOT}}(\pi(F_n,G_m)).
\end{equation}
The next result is the key for the performance of 
$\hat{\pi}_{n,m,\mbox{\scriptsize BOOT}}$.
\begin{Prop}\label{vanishingbias} Under the assumptions of Theorem \ref{Ragha}
we have
$${\textstyle \sqrt{\frac{mn}{m+n}}}\widehat{\mbox{\em bias}}_{\mbox{\scriptsize BOOT}}(\pi(F_n,G_m))\to 0.$$
in probability as $n,m\to\infty$.
\end{Prop}

A proof is given in the Appendix. Proposition \ref{vanishingbias} shows that $\sqrt{\frac{mn}{m+n}} (
\pi(F_n^*,G_m^*)-\pi(F_n,G_m))$ does not mimick the asymptotic behavior of $\sqrt{\frac{mn}{m+n}} (
\pi(F_n,G_m)-\pi(F,G))$ unless $\pi(F_n,G_m)$ is asymptotically normal. But, more importantly, it shows that
the bootstrap bias correction does not affect the first order behavior of $\pi(F_n,G_m)$. In other words,
that rejection of 
$H_0:\, \pi(F,G)\geq \pi_0$ if
\begin{equation}\label{proposal3}
\textstyle{\sqrt{\frac{nm}{n+m}}
(\hat{\pi}_{n,m,\mbox{\scriptsize BOOT}}-\pi_0) < \hat{\sigma}_{n,m}\Phi^{-1}(\alpha)},
\end{equation}
with $\hat{\sigma}_{n,m}$ as above, is a test of asymptotic level $\alpha$ with fastly
decreasing type I and type II error probabilities away from the null hypothesis boundary.
We show later, in a simulation study in Section 4, that the bootstrap correction (which 
of course has to be approximated, in turn, through bootstrap simulation) yields a significant
improvement with respect to (\ref{proposal1}) or (\ref{proposal2}) in terms of power and 
approximation of the nominal level for finite samples.

\subsection{Testing against essential stochastic order}

In some instances the researcher can be interested in gathering statistical
evidence against stochastic order, or to stochastic order up to some small
contamination. The relevant testing problem is then (\ref{nullmodelbis}), namely,
\begin{equation}\label{modelagainst}
H_0:\, \pi(F,G)\leq \pi_0
\end{equation}
against the alternative $H_a:\, \pi(F,G)> \pi_0$.
Now, we would reject the null hypothesis for large values of $\pi(F_n,G_m)$.
Motivated by Proposition \ref{quantilebounds} we consider the test that rejects
$H_0$ in (\ref{modelagainst}) if 
\begin{equation}\label{proposalagainst}
\textstyle{\sqrt{\frac{nm}{n+m}}} 
(\pi_{n,m}-\pi_0) > K_{1-\alpha}(\pi_0,\lambda),
\end{equation}
where $K_{1-\alpha}(\pi_0,\lambda)$ is the $1-\alpha$ quantile of $\bar{B}(\pi_0,\lambda)$
defined in (\ref{alternativeextra}). Next, we give the main facts about the test (\ref{proposalagainst}).
As in the statement of Proposition \ref{Test1} we 
write $\pi_{m,n}$ for $\pi(F_n,G_m)$ and $\mathbb{P}_{F,G}$ for the probabilities under the
assumption that the underlying distribution functions of the two samples are $F$ and $G$, respectively.

\begin{Prop}\label{TestAgainst}
With the above assumptions and notation,
\begin{eqnarray}\nonumber
\lefteqn{\lim_{n\to\infty} \sup_{(F,G)\in H_0} \mathbb{P}_{F,G}(  \textstyle{\sqrt{\frac{nm}{n+m}}
(\pi_{n,m}-\pi_0) > K_{1-\alpha}(\pi_0,\lambda)} )}\hspace*{1cm}\\
\label{asymptoticlevelag}
&=& \lim_{n\to\infty}  \mathbb{P}_{F_0,G_0}(  \textstyle{\sqrt{\frac{nm}{n+m}}
(\pi_{n,m}-\pi_0)> K_{1-\alpha}(\pi_0,\lambda)} ) =\alpha,
\end{eqnarray}
where $F_0$ is the distribution function of the law 
$U(\pi_0,1+\pi_0)$  and
$G_0$ is the distribution function of the law $U(0,1)$. Furthermore, if $\pi(F,G)<\pi_0$ and $ K_{1-\alpha}(\pi_0,\lambda)\geq 0$
then
\begin{equation}\label{errortipo1ag}
\mathbb{P}_{F,G}(\textstyle{\sqrt{\frac{nm}{n+m}}
(\pi_{n,m}-\pi_0) > K_{1-\alpha}(\pi_0,\lambda)})
\leq 2 e^{-\frac{2}{1+2\sqrt{\lambda_n(1-\lambda_n)}}
\frac{nm}{n+m}(\pi_0-\pi(F,G))^2},
\end{equation}
while if $\pi(F,G)>\pi_0$ then 
\begin{equation}\label{errortipo2ag}
\mathbb{P}_{F,G}(\textstyle{\sqrt{\frac{nm}{n+m}}
(\pi_{n,m}-\pi_0) \leq K_{1-\alpha}(\pi_0,\lambda)})
\leq e^{-2(\sqrt{\frac{nm}{n+m}}(\pi_0-\pi(F,G)) - K_{1-\alpha}(\pi_0,\lambda))^2}.
\end{equation}
\end{Prop}
A proof of Proposition \ref{TestAgainst} is given in the Appendix. 
Similar comments as in Remark \ref{expunifconsistent} can be made now. 
The test in (\ref{proposalagainst}) is asymptotically
of level $\alpha$ for $H_0: \pi(F,G)\leq \pi_0$ vs. $H_a: \pi(F,G)>\pi_0$
and uniformly exponentially consistent test for $H'_0: \pi(F,G)\leq \pi_0'$
vs. $H_a': \pi(F,G)>\pi_1$ if $\pi_0'<\pi_0<\pi_1$. Later, in Section 4
we will see that this test shows a good performance for finite sample
sizes (even for small sizes).

\subsection{Confidence bounds.}

Rather than testing for or against the contaminated stochastic order model one
could prefer to report results in terms of confidence intervals for the true
contamination level, $\pi(F,G)$. Here we discuss briefly upper and lower confidence
bounds for $\pi(F,G)$. Proper two-sided confidence intervals can we constructed
from our confidence bounds in a straighforward way. We omit details.

\medskip
Recalling Theorem \ref{Ragha} and Proposition \ref{quantilebounds} we see that
$$\pi(F_n,G_m)-\textstyle \sqrt{\frac{n+m}{nm}}K_\alpha(F,G,\lambda),$$
is an ideal upper confidence bound, asymptotically of level $1-\alpha$ for $\pi(F,G)$, that 
cannot be used directly, since the quantiles $K_\alpha(F,G,\lambda)$ are unknown.
It follows from Theorem \ref{Ragha}
that $K_\alpha(F,G,\lambda)$ can be consistently estimated by the conditional $\alpha$-quantile
of $ \sqrt \frac {mn}{m+n}\sup_{x\in \Gamma_n(F_n,G_m)}((G_m^*(x)-G_m(x))-(F_n^*(x)-F_n(x))))$, 
that we denote by $\hat{K}_\alpha^{\mbox{\scriptsize (Boot)}}$ which can be approximated by
simulation. As a result, we have that
\begin{equation}\label{upperconfidence}
\pi(F_n,G_m)-\textstyle \sqrt{\frac{n+m}{nm}} \hat{K}_\alpha^{\mbox{\scriptsize (Boot)}}
\end{equation}
is an upper confidence bound for $\pi(F,G)$ with asymptotic confidence level $1-\alpha$.  
Unfortunately, as we can see in Table ??? below, the
finite sample performance of this upper confidence bound or test can be too liberal even for 
large sample sizes. Hence it could be better to consider different confidence bounds.

Assuming $\alpha<\frac 1 2$, it follows from Propositions \ref{quantilebounds} and
\ref{vanishingbias}
that 
\begin{equation}\label{upperconfidencedirect}
\hat{\pi}_{n,m,\mbox{\scriptsize BOOT}}-\textstyle \sqrt{\frac{n+m}{nm}} \hat{\sigma}_{m,n}\Phi^{-1}(\alpha)
\end{equation}
with $\hat{\pi}_{n,m,\mbox{\scriptsize BOOT}}$ and $\hat{\sigma}_{m,n}$ as in
(\ref{proposal3}) is an upper bound with asymptotic confidence level at least $1-\alpha$. 
Our simulations in Section 4 show a good performance of (\ref{upperconfidencedirect})
for finite samples.

Turning to the issue of lower confidence bounds, Theorem \ref{Ragha} and Proposition \ref{quantilebounds}
imply that
\begin{equation}\label{lowerconfidence}
\pi(F_n,G_m)-\textstyle \sqrt{\frac{n+m}{nm}} K_{1-\alpha}(\pi(F_n,G_m),\lambda_{m,n})
\end{equation}
is a lower confidence bound for $\pi(F,G)$ with asymptotic confidence level $1-\alpha$. 
As for the test in (\ref{proposalagainst}),
quantiles $K_{1-\alpha}(\pi(F_n,G_m),\lambda_{m,n})$ can be numerically approximated 
from part (a) of Proposition \ref{DistRepre}.

\section{Simulations and Case Study}

We explore here the finite sample performance of the tests and confidence bounds introduced
in Section 3. We start with the tests for essential stochastic order (\ref{proposal1}),
(\ref{proposal2}) and (\ref{proposal3}). We consider several values of $\pi_0$
and have simulated from different pairs $(F,G)$.
Proposition \ref{Test1} tells us that (at least asymptotically) for a fixed value 
of $\pi=\pi(F,G)$, type I error probability is largest for $F_{\pi,b}$ 
corresponding to $(\frac 1 2-\pi\lambda)U(0,\frac 1 2+\pi(1-\lambda))+ 
(\frac 1 2 +\pi\lambda)U(\frac 1 2+\pi(1-\lambda),1+\frac {\pi} 2-\lambda \pi^2)$ 
and $G$ coming from the uniform law on $(0,1)$, while from the point of view of power the
worst performance (recall Theorem \ref{Ragha} and Proposition \ref{quantilebounds}) should
come from the pair $(F_{\pi,a},G)$ with $F_{\pi,a}$ the d.f. of the uniform law on $(\pi,1+\pi)$
and $G$ as before. Consequently, we have simulated samples from these choices $F_{\pi,a}$,
$F_{\pi,b}$ and $G$ for several values of $\pi$. We have also considered the case $F_0=G$. 
Although we have some indication
that the balance of sample sizes has an impact on the performance of the procedure
(see the comments after Proposition \ref{DistRepre}) we have, for the sake of brevity,
focused on the case $m=n$ and have considered different sample sizes. In the next tables
we show the simulated rejection frequencies for the tests (\ref{proposal1}),
(\ref{proposal2}) and (\ref{proposal3}). In all cases we have computed this simulated
rejection frequency from 1000 replicates of the procedure. In the case of test 
(\ref{proposal3}) the bootstrap bias correction has been approximated by the average
from 1000 bootstrap replicates. In all cases the nominal level of the test was 
$\alpha=0.05$ and $G$ is the d.f. of the uniform law on $(0,1)$.

\begin{table}[ht]
\caption{\label{potencia1}Observed rejection frequencies.
$H_0: \pi(F,G)\geq \pi_0$ vs. $H_a: \pi(F,G)<\pi_0$ \newline
\hspace*{1.5cm} $G=U(0,1)$, $m=n$; reject if ${\sqrt{\frac{nm}{n+m}}
(\pi_{n,m}-\pi_0) < \bar{\sigma}_{\pi_0}\Phi^{-1}(0.05)}$}

\centering
\begin{footnotesize}
\begin{tabular}{|c|r|rr|rr|rr|rr|r|}
\hline
\multicolumn{1}{|c}{\boldmath{$\pi_0$}}&\multicolumn{1}{|c|}{\boldmath{$n$}} & \boldmath{$F_{0.2,a}$} &\boldmath{$F_{0.2,b}$}&
\boldmath{$F_{0.1,a}$}&\boldmath{$F_{0.1,b}$}&\boldmath{$F_{0.05,a}$}&\boldmath{$F_{0.05,b}$} &\boldmath{$F_{0.01,a}$} &\boldmath{$F_{0.01,b}$} 
&\boldmath{$F_{0}$}\\
 \hline
\multirow{6}{10mm}{\centering\textbf{0.01}}
                              &50   &  0.000  &0.000&0.000 &0.000 &0.000&0.000 &0.000&0.000&0.000\\
                              &100  &  0.000  &0.000&0.000 &0.000 &0.000&0.000 &0.000&0.000&0.000\\
                              &500  &  0.000  &0.000&0.000 &0.000 &0.000&0.000 &0.000&0.000&0.000\\
                              &1000 &  0.000  &0.000&0.000 &0.000 &0.000&0.000 &0.000&0.000&0.000\\
                              &5000 &  0.000  &0.000&0.000 &0.000 &0.000&0.000 &0.000&0.000&0.000\\
                              &10000&  0.000  &0.000&0.000 &0.000 &0.000&0.000 &0.000&0.000&0.000\\
\hline                                 
\multirow{6}{10mm}{\centering\textbf{0.05}}
                              &50   &0.000&0.000&0.000&0.000&0.000& 0.000 & 0.000 & 0.000 &0.000\\
                              &100  &0.000&0.000&0.000&0.000&0.000& 0.000 & 0.000 & 0.000 &0.000\\
                              &500  &0.000&0.000&0.000&0.000&0.000& 0.000 & 0.000 & 0.000 &0.000\\
                              &1000 &0.000&0.000&0.000&0.000&0.000& 0.000 & 0.016 & 0.071 &0.183\\
                              &5000 &0.000&0.000&0.000&0.000&0.000& 0.008 & 0.924 & 0.939 &0.995\\
                              &10000&0.000&0.000&0.000&0.000&0.000& 0.021 & 0.999 & 1.000 &1.000\\
\hline                                
\multirow{6}{10mm}{\centering\textbf{0.1}}
                              &50   &0.000&0.000&0.000&0.000&0.000&0.000&0.000&0.000&0.000\\
                              &100  &0.000&0.000&0.000&0.000&0.000&0.000&0.000&0.000&0.000\\
                              &500  &0.000&0.000&0.000&0.001&0.010&0.152&0.532&0.589&0.698\\
                              &1000 &0.000&0.000&0.000&0.011&0.155&0.491&0.946&0.952&0.979\\
                              &5000 &0.000&0.000&0.000&0.025&0.996&1.000&1.000&1.000&1.000\\
                              &10000&0.000&0.000&0.000&0.036&1.000&1.000&1.000&1.000&1.000\\
                              \hline      
\multirow{6}{10mm}{\centering\textbf{0.2}}
                              &50   &0.000&0.000&0.000&0.004&0.006&0.024&0.045&0.048&0.075\\
                              &100  &0.000&0.002&0.010&0.087&0.151&0.296&0.480&0.499&0.559\\
                              &500  &0.000&0.021&0.706&0.880&0.997&0.996&1.000&1.000&1.000\\
                              &1000 &0.000&0.022&0.985&0.994&1.000&1.000&1.000&1.000&1.000\\
                              &5000 &0.000&0.047&1.000&1.000&1.000&1.000&1.000&1.000&1.000\\
                              &10000&0.000&0.041&1.000&1.000&1.000&1.000&1.000&1.000&1.000\\
\hline                                 
\end{tabular}
\end{footnotesize}
\end{table}

\begin{table}[ht]
\caption{\label{potencia2}Observed rejection frequencies.
$H_0: \pi(F,G)\geq \pi_0$ vs. $H_a: \pi(F,G)<\pi_0$ \newline
\hspace*{1.5cm} $G=U(0,1)$, $m=n$; reject if ${\sqrt{\frac{nm}{n+m}}
(\pi_{n,m}-\pi_0) < \hat{\sigma}_{m,n}\Phi^{-1}(0.05)}$}

\centering
\begin{footnotesize}
\begin{tabular}{|c|r|rr|rr|rr|rr|r|}
\hline
\multicolumn{1}{|c}{\boldmath{$\pi_0$}}&\multicolumn{1}{|c|}{\boldmath{$n$}} & \boldmath{$F_{0.2,a}$} &\boldmath{$F_{0.2,b}$}&
\boldmath{$F_{0.1,a}$}&\boldmath{$F_{0.1,b}$}&\boldmath{$F_{0.05,a}$}&\boldmath{$F_{0.05,b}$} &\boldmath{$F_{0.01,a}$} &\boldmath{$F_{0.01,b}$} 
&\boldmath{$F_{0}$}\\
 \hline
\multirow{6}{10mm}{\centering\textbf{0.01}}
                              &50   &0.000&0.000&0.000&0.000&0.000&0.003&0.007&0.015&0.017\\
                              &100  &0.000&0.000&0.000&0.000&0.000&0.000&0.000&0.005&0.016\\
                              &500  &0.000&0.000&0.000&0.000&0.000&0.000&0.000&0.001&0.007\\
                              &1000 &0.000&0.000&0.000&0.000&0.000&0.000&0.000&0.003&0.016\\
                              &5000 &0.000&0.000&0.000&0.000&0.000&0.000&0.000&0.012&0.051\\
                              &10000&0.000&0.000&0.000&0.000&0.000&0.000&0.000&0.023&0.117\\
\hline                                 
\multirow{6}{10mm}{\centering\textbf{0.05}}
                              &50   &0.000&0.000&0.000&0.001&0.000&0.003&0.005&0.011&0.017\\
                              &100  &0.000&0.000&0.000&0.001&0.000&0.007&0.020&0.036&0.039\\
                              &500  &0.000&0.000&0.000&0.000&0.000&0.015&0.100&0.105&0.147\\
                              &1000 &0.000&0.000&0.000&0.000&0.000&0.022&0.206&0.231&0.360\\
                              &5000 &0.000&0.000&0.000&0.000&0.000&0.023&0.948&0.973&0.998\\
                              &10000&0.000&0.000&0.000&0.000&0.000&0.025&1.000&1.000&1.000\\
\hline                                
\multirow{6}{10mm}{\centering\textbf{0.1}}
                              &50   &0.000&0.000&0.000&0.011&0.010&0.042&0.063&0.064&0.080\\
                              &100  &0.000&0.000&0.000&0.011&0.019&0.033&0.123&0.124&0.137\\
                              &500  &0.000&0.000&0.000&0.017&0.151&0.216&0.606&0.693&0.761\\
                              &1000 &0.000&0.000&0.000&0.025&0.337&0.523&0.961&0.957&0.979\\
                              &5000 &0.000&0.000&0.000&0.028&0.993&0.999&1.000&1.000&1.000\\
                              &10000&0.000&0.000&0.000&0.033&1.000&1.000&1.000&1.000&1.000\\
                              \hline      
\multirow{6}{10mm}{\centering\textbf{0.2}}
                              &50   &0.000& 0.015& 0.048& 0.101&0.151&0.171&0.273&0.295&0.288\\
                              &100  &0.000& 0.013& 0.118& 0.175&0.369&0.406&0.580&0.614&0.643\\
                              &500  &0.000& 0.020& 0.770& 0.880&0.993&0.993&1.000&0.999&1.000\\
                              &1000 &0.000& 0.033& 0.988& 0.995&1.000&1.000&1.000&1.000&1.000\\
                              &5000 &0.000& 0.038& 1.000& 1.000&1.000&1.000&1.000&1.000&1.000\\
                              &10000&0.000& 0.032& 1.000& 1.000&1.000&1.000&1.000&1.000&1.000\\
\hline                                    
\end{tabular}
\end{footnotesize}
\end{table}

\begin{table}[ht]
\caption{\label{potencia3}Observed rejection frequencies.
$H_0: \pi(F,G)\geq \pi_0$ vs. $H_a: \pi(F,G)<\pi_0$ \newline
\hspace*{1.5cm} $G=U(0,1)$, $m=n$; reject if ${\sqrt{\frac{nm}{n+m}}
(\hat{\pi}_{n,m,\mbox{\scriptsize BOOT}}-\pi_0) < \hat{\sigma}_{n,m}\Phi^{-1}(0.05)}$}

\centering
\begin{footnotesize}
\begin{tabular}{|c|r|rr|rr|rr|rr|r|}
\hline
\multicolumn{1}{|c}{\boldmath{$\pi_0$}}&\multicolumn{1}{|c|}{\boldmath{$n$}} & \boldmath{$F_{0.2,a}$} &\boldmath{$F_{0.2,b}$}&
\boldmath{$F_{0.1,a}$}&\boldmath{$F_{0.1,b}$}&\boldmath{$F_{0.05,a}$}&\boldmath{$F_{0.05,b}$} &\boldmath{$F_{0.01,a}$} &\boldmath{$F_{0.01,b}$} 
&\boldmath{$F_{0}$}\\
 \hline
\multirow{6}{10mm}{\centering\textbf{0.01}}
                              &50   &0.000&0.000&0.000&0.003&0.001&0.012&0.019&0.034&0.038\\
                              &100  &0.000&0.000&0.000&0.001&0.000&0.005&0.015&0.027&0.028\\
                              &500  &0.000&0.000&0.000&0.000&0.000&0.001&0.016&0.028&0.049\\
                              &1000 &0.000&0.000&0.000&0.000&0.000&0.000&0.013&0.030&0.063\\
                              &5000 &0.000&0.000&0.000&0.000&0.000&0.000&0.019&0.038&0.136\\
                              &10000&0.000&0.000&0.000&0.000&0.000&0.000&0.013&0.051&0.277\\
\hline                  
\multirow{6}{10mm}{\centering\textbf{0.05}}
                              &50   &0.000&0.000&0.000&0.009&0.015&0.042&0.062&0.067&0.093\\
                              &100  &0.000&0.000&0.000&0.006&0.008&0.038&0.065&0.098&0.106\\
                              &500  &0.000&0.000&0.000&0.003&0.007&0.058&0.208&0.220&0.332\\
                              &1000 &0.000&0.000&0.000&0.001&0.009&0.039&0.415&0.426&0.566\\
                              &5000 &0.000&0.000&0.000&0.000&0.003&0.057&0.978&0.987&1.000\\
                              &10000&0.000&0.000&0.000&0.000&0.006&0.050&1.000&1.000&1.000\\
\hline                                
\multirow{6}{10mm}{\centering\textbf{0.1}}
                              &50   &0.000&0.005&0.003&0.030&0.054&0.089&0.137&0.138&0.134\\
                              &100  &0.000&0.001&0.007&0.052&0.076&0.121&0.246&0.250&0.266\\
                              &500  &0.000&0.000&0.007&0.040&0.337&0.387&0.801&0.830&0.876\\
                              &1000 &0.000&0.000&0.005&0.056&0.589&0.661&0.976&0.985&0.997\\
                              &5000 &0.000&0.000&0.003&0.057&0.999&1.000&1.000&1.000&1.000\\
                              &10000&0.000&0.000&0.008&0.058&1.000&1.000&1.000&1.000&1.000\\
                              \hline      
\multirow{6}{10mm}{\centering\textbf{0.2}}
                              &50   &0.007&0.051&0.130&0.172&0.321&0.350&0.483&0.469&0.508\\
                              &100  &0.003&0.068&0.280&0.339&0.541&0.600&0.758&0.761&0.798\\
                              &500  &0.004&0.051&0.888&0.928&1.000&0.997&1.000&1.000&1.000\\
                              &1000 &0.002&0.050&0.999&0.999&1.000&1.000&1.000&1.000&1.000\\
                              &5000 &0.002&0.054&1.000&1.000&1.000&1.000&1.000&1.000&1.000\\
                              &10000&0.004&0.045&1.000&1.000&1.000&1.000&1.000&1.000&1.000\\
\hline                                    
\end{tabular}
\end{footnotesize}
\end{table}

We see in Table \ref{potencia1} 
how alternatives are detected with power rapidly increasing to 1, as predicted by
(\ref{power}). For instance, in this balanced setup ($m=n$), if we fix $\pi_0=0.05$
(we are trying to establish that $F$ is stochastically smaller than $G$ up to 5\% contamination)
then, to guarantee that alternatives with $\pi(F,G)=0.1$ are detected with power
at least 90\% the bound (\ref{power}) requires a sample size $n=m=8143$. In the 
simulation study we observe that the power is above 90\% for $n=m=5000$.
We also see the very small type I error probability guaranteed by (\ref{errortipo1}).
In fact, we see that the test in (\ref{proposal1}) is somewhat conservative for
finite samples with slow convergence to the nominal level This is more clearly seen 
for small values of $\pi_0$. In Table \ref{potencia2} we see that the correction (\ref{proposal2})
improves slightly the convergence to the nominal level, resulting in some
increase in power while keeping the low type I error probabilities. Table \ref{potencia3} 
shows the remarkable effect of the bootstrap bias correction (\ref{proposal3}).
We see that sample sizes about $n=m=1000$ suffice to give a rather close agreement
to the nominal level, even for small values of $\pi_0$. And we also see that the bias
correction results in a significant increase in power. As an example, if we are
trying to reject that there is more than 10\% contamination with respect to the
stochastic order model and we were, in fact, sampling from distributions with 5\%
contamination or less, then samples of size 1000 would give a probability of rejection of
$60\%$ or more and from samples of size 5000 we would reject with probability close to 1.
Even for the hard problem of concluding that $F$ and $G$ satisfy the stochastic order up
to $1\%$ contamination we see nonnegligible power for $n=5000$ or 10000, a sample size
not unusual in some econometric studies (for instance, the Canadian Family Expenditure Survey,
considered in \cite{Barrett} involves more than 9000 units).

\bigskip
In the testing problem (\ref{modelagainst}), namely, the problem of looking
for statistcal evidence against stochastic order up to some small contamination
we have considered the test (\ref{proposalagainst}), that is, 
$H_0: \pi(F,G) \leq \pi_0$ is rejected if 
$\sqrt{\frac n 2}(\pi(F_n,G_n)-\pi_0)> K_{1-\alpha}(\pi_0,\frac 1 2)$.
$K_{1-\alpha}$ has been aproximated using the expresion in Proposition \ref{DistRepre}
(a) plus numerical integration and inversion. As before, we have focused on the case $m=n$ and 
$\alpha=0.05$. In this case, for a fixed value of $\pi=\pi(F,G)$, the worst case from
the point of view of type I error corresponds to $\tilde{F}_{\pi,a}=U(\pi,1+\pi)$ vs $G=U(0,1)$
while the worst case for power is $\tilde{F}_{\pi,b}=
\frac{1-\pi}2 U(0,\frac{1+\pi}2)+\frac{1+\pi}2U(\frac{1+\pi}2,1+\frac{\pi(1-\pi)}2)$
vs. $G=U(0,1)$ and these are the distributions that we have considered. 
Again, we have also considered $\tilde{F}_0=U(0,1)$. The results are reported in Table
\ref{potenciaagainst}. We observe a very good agreement between nominal and simulated
levels, even for small values of $n$. We also see rapidly decaying error probabilites
as predicted by (\ref{errortipo1ag}) and (\ref{errortipo2ag}). 

\begin{table}[ht]
\caption{\label{potenciaagainst}Observed rejection frequencies.
$H_0: \pi(F,G)\leq \pi_0$ vs. $H_a: \pi(F,G)>\pi_0$ \newline
\hspace*{1.5cm} $G=U(0,1)$, $m=n$; reject if ${\sqrt{\frac{nm}{n+m}}
({\pi}_{n,m}-\pi_0) > K_{0.95}(\pi_0,\lambda_{n,m})}$}

\centering
\begin{footnotesize}
\begin{tabular}{|c|r|r|rr|rr|rr|rr|}
\hline
\multicolumn{1}{|c}{\boldmath{$\pi_0$}}&\multicolumn{1}{|c|}{\boldmath{$n$}} & \boldmath{$\tilde{F}_{0}$} &\boldmath{$\tilde{F}_{0.01,a}$}&
\boldmath{$\tilde{F}_{0.01,b}$}&\boldmath{$\tilde{F}_{0.05,a}$}&\boldmath{$\tilde{F}_{0.05,b}$}&\boldmath{$\tilde{F}_{0.1,a}$} &\boldmath{$\tilde{F}_{0.1,b}$} &
\boldmath{$\tilde{F}_{0.2,a}$} 
&\boldmath{$\tilde{F}_{0.2,b}$}\\
 \hline
\multirow{6}{10mm}{\centering\textbf{0.01}}
                              &50   &0.045&0.039&0.052&0.060&0.111&0.159&0.302&0.485&0.824\\
                              &100  &0.022&0.031&0.040&0.066&0.107&0.199&0.450&0.848&0.986\\
                              &500  &0.021&0.026&0.045&0.210&0.477&0.951&1.000&1.000&1.000\\
                              &1000 &0.011&0.028&0.047&0.441&0.774&1.000&1.000&1.000&1.000\\
                              &5000 &0.002&0.017&0.049&1.000&1.000&1.000&1.000&1.000&1.000\\
                              &10000&0.001&0.005&0.047&1.000&1.000&1.000&1.000&1.000&1.000\\
\hline                  
\multirow{6}{10mm}{\centering\textbf{0.05}}
                              &50   &0.015&0.010&0.016&0.023&0.052&0.056&0.142&0.253&0.600\\
                              &100  &0.004&0.007&0.009&0.014&0.031&0.069&0.186&0.518&0.885\\
                              &500  &0.000&0.001&0.002&0.009&0.047&0.211&0.664&1.000&1.000\\
                              &1000 &0.000&0.000&0.000&0.001&0.060&0.606&0.954&1.000&1.000\\
                              &5000 &0.000&0.000&0.000&0.001&0.040&1.000&1.000&1.000&1.000\\
                              &10000&0.000&0.000&0.000&0.000&0.056&1.000&1.000&1.000&1.000\\
\hline                                
\multirow{6}{10mm}{\centering\textbf{0.1}}
                              &50   &0.001&0.002&0.005&0.004&0.007&0.009&0.027&0.079&0.274\\
                              &100  &0.000&0.003&0.000&0.001&0.002&0.010&0.031&0.163&0.520\\
                              &500  &0.000&0.000&0.000&0.000&0.001&0.002&0.056&0.958&0.999\\
                              &1000 &0.000&0.000&0.000&0.000&0.000&0.001&0.035&1.000&1.000\\
                              &5000 &0.000&0.000&0.000&0.000&0.000&0.000&0.056&1.000&1.000\\
                              &10000&0.000&0.000&0.000&0.000&0.000&0.000&0.057&1.000&1.000\\
                              \hline      
\multirow{6}{10mm}{\centering\textbf{0.2}}
                              &50   &0.000&0.000&0.000&0.000&0.000&0.001&0.004&0.004&0.029\\
                              &100  &0.000&0.000&0.000&0.000&0.000&0.000&0.000&0.002&0.051\\
                              &500  &0.000&0.000&0.000&0.000&0.000&0.000&0.000&0.001&0.044\\
                              &1000 &0.000&0.000&0.000&0.000&0.000&0.000&0.000&0.001&0.052\\
                              &5000 &0.000&0.000&0.000&0.000&0.000&0.000&0.000&0.000&0.053\\
                              &10000&0.000&0.000&0.000&0.000&0.000&0.000&0.000&0.000&0.040\\
\hline                                    
\end{tabular}
\end{footnotesize}
\end{table}

\section{Discussion}

In the stochastic dominance setting, the approximation to the model 
given by the mixture approach can be easily characterized and strongly 
suggests the estimation of the mixture index as well as the testing 
statistics. In fact this mixture index gives a nice interpretation to 
the meaning of the one-sided Kolmogorov-Smirnov statistic.This is a 
consequence of the fact that although the contaminating distributions 
can distort the model in very different ways, there exist a most unfavorable 
way for that, thus the less favorable hypothesis can be stated in terms of 
just two distributions involving the parent distributions $P_1$ and $P_2$ 
and the maximum level of contamination allowed, say $\pi_0$.

\section*{Appendix}

In this Appendix we provide proofs for the results in Section 3. Most 
of them are related to the behavior of $\pi(F_n,G_m)$. We keep the notation 
of Section 3, including that for the sets
$$\Gamma(F,G)=\{x\in\bar{\mathbb{R}}: G(x)-F(x)=\pi(F,G)\}$$
and
$$T(F,G,\pi)=\{t\in[0,1]: G(x)=t, F(x)=t-\pi \mbox{ for some }x\in\bar{\mathbb{R}}\}$$
(we write $\bar{\mathbb{R}}=\mathbb{R}\cup \{-\infty,+\infty \}$ in the definition of the sets,
with $G(+\infty)-F(+\infty)=G(-\infty)-F(-\infty)=0$,
to cover the case $\pi(F,G)=0$ with $G(x)<F(x)$ for all $x\in \mathbb{R}$).

Throughout this Appendix we will assume (without loss of generality) 
that $X_i=F^{-1}(U_i)$, $i=1,\ldots,n$ and $Y_j=G^{-1}(V_j)$, $j=1,\ldots,m$
where $U_1,\ldots,U_n,V_1,\ldots,V_m$ are i.i.d. r.v.'s. We will write 
$\alpha_{m,1}(t)$, $0\leq t \leq 1$ for the empirical process based on the $V_j$'s
and $\alpha_{n,2}$ for the empirical process on the $U_i$'s. We note that, in particular,
$G_m(x)=G(x)+\frac{1}{\sqrt{m}}\alpha_{m,1}(G(t))$ and 
$F_n(x)=F(x)+\frac{1}{\sqrt{n}}\alpha_{n,2}(F(t))$. We will use this fact throughout this 
Appendix without further mention. 
We introduce the processes
\begin{eqnarray}\label{twosampleempirical}
\alpha_{m,n}(s,t)&=&\sqrt{\lambda_{n,m}}\alpha_{m,1}(s)-\sqrt{1-\lambda_{n,m}}
\alpha_{n,2}(t),\quad 0\leq s,t\leq 1,\\
B_\lambda(s,t)&=&\sqrt{\lambda}B_1(s)-\sqrt{1-\lambda}B_2(t),\quad 0\leq s,t\leq 1,
\end{eqnarray}
where $\lambda_{n,m}=\frac{n}{n+m}\to \lambda\in(0,1)$ and $B_1,B_2$ are independent Brownian
bridges on $[0,1]$.
Finally, we will write $\|\cdot\|_\infty$ for the sup norm
in $[0,1]$ and $\omega_{m,1}(\delta),\omega_{n,2}(\delta)$ for the oscillation modulus 
of the empirical processes $\alpha_{m,1}$ and $\alpha_{n,2}$, respectively, namely,
$$\omega_{m,1}(\delta)=\sup_{0\leq t-s\leq \delta} |\alpha_{m,1}(s)-\alpha_{m,1}(t)|$$
and similarly for $\omega_{n,2}$.

\medskip
The following estimates give the key to the asymptotic distributional 
behavior of the estimator $\pi(F_n,G_m)$.
\begin{Lemm}\label{upperlowerbound}
If we denote $\Delta_{n,m}=2(m^{-1/2}\|\alpha_{m,1}\|_\infty+n^{-1/2}\|\alpha_{n,2}\|_\infty)$,
$\tilde{\Gamma}_{\delta}(F,G)=\{x\in\mathbb{R}:\, G(x)-F(x)\geq \pi(F,G)-\delta\}$
and
$$R_{n,m}=\sqrt{\lambda_{n,m}}\omega_{m,1}(\Delta_{n,m})+
\sqrt{1-\lambda_{n,m}}\omega_{n,2}(\Delta_{n,m})$$ then
\begin{eqnarray}\nonumber
\sup_{x\in \Gamma(F,G)}
\alpha_{n,m}(G(x),F(x))&\leq &\textstyle{\sqrt{\frac{mn}{m+n}} }(\pi(F_n,G_m)-\pi(F,G))\\
\label{mainupperlowerbound} &\leq &
\sup_{x\in \tilde{\Gamma}_{\Delta_{n,m}}(F,G)}\alpha_{n,m}(G(x),F(x))
\end{eqnarray}
and
\begin{equation}\label{uppercontrol}
{\textstyle\sqrt{\frac{mn}{m+n}} }(\pi(F_n,G_m)-\pi(F,G))\leq \sup_{t\in [\pi(F,G),1]} \alpha_{n,m}(t,t-\pi(F,G)) +R_{n,m}.
\end{equation}
\end{Lemm}

\medskip
\noindent
\textbf{Proof.} 
We recall that $\Gamma(F,G):=\{x\in\bar{\mathbb{R}}: 
G(x)-F(x)=\pi(F,G) \}$. Hence, if $x\in\Gamma(F,G)$ then $G_m(x)-F_n(x)=(G_m(x)-G(x))-(F_n(x)-F(x))+
\pi(F,G)=\sqrt{\frac{n+m}{nm}}\alpha_{n,m,\pi(F,G)}(G(x))+\pi(F,G)$. From this obtain
the lower bound in (\ref{mainupperlowerbound}).
Also, writing $G_m(x)-F_n(x)=(G_m(x)-G(x))-(F_n(x)-F(x))+(G(x)-F(x))$ we see that
$G_m(x)-F_n(x)\leq G(x)-F(x)+m^{-1/2}\|\alpha_{m,1}\|_\infty$ 
$+n^{-1/2}\|\alpha_{n,2}\|_\infty$ while for any $x\in\Gamma(F,G)$,
$G_m(x)-F_n(x)\geq \pi(F,G)-m^{-1/2}\|\alpha_{m,1}\|_\infty$ 
$-n^{-1/2}\|\alpha_{n,2}\|_\infty$.
Therefore, for any $x'$ outside
$\tilde{\Gamma}_{\Delta_{n,m}}(F,G)$ and any $x\in\Gamma(F,G)$
$$G_m(x')-F_n(x')<\pi(F,G)-m^{-1/2}\|\alpha_{m,1}\|_\infty+n^{-1/2}\|\alpha_{n,2}\|_\infty \leq G_m(x)-F_n(x),$$
which means that $\pi(F_n,G_m)=\sup_{x\in\tilde{\Gamma}_{\Delta_{n,m} }(F,G)}(G_m(x)-F_n(x))$.
As a consequence 
$${\textstyle\sqrt{\frac{mn}{m+n}} }(\pi(F_n,G_m)-\pi(F,G))\leq \sup_{
x\in\tilde{\Gamma}_{\Delta_{n,m} }(F,G)} \sqrt{\lambda_{n,m}} 
\alpha_{m,1}(G(x))-\sqrt{1-\lambda_{n,m}} \alpha_{n,2}(F(x)),$$
giving the upper bound in (\ref{mainupperlowerbound}). 

Consider now $x\in \tilde{\Gamma}_{\Delta_{n,m} }(F,G)$. If $G(x)\leq\pi(F,G)$ then
$G(x)\geq\pi(F,G)-\Delta_{n,m}$ and $F(x)\leq \Delta_{m,n}$, from which
we see that
$$\alpha_{m,n}(G(x),F(x))\leq
\alpha_{n,m}(\pi(F,G),0)+R_{n,m}.$$
On the other hand, if $x\in \tilde{\Gamma}_{\Delta_{n,m} }(F,G)$ and $G(x)\geq\pi(F,G)$
then $G(x)-\pi(F,G)+\delta_{n,m}\geq F(x) \geq G(x)-\pi(F,G)\geq 0$ and this entails
$$\alpha_{m,n}(G(x),F(x))\leq
\alpha_{n,m}(G(x),G(x)-\pi(F,G))+R_{n,m}.$$
Combining the last two estimates we conclude (\ref{uppercontrol}).
\hfill $\Box$

\medskip

\medskip
\noindent
\textbf{Proof of Theorem \ref{Ragha}.} 
From (\ref{mainupperlowerbound}) we see that the result will follow if we show
that 
\begin{equation}\label{convlower}
\sup_{x\in\Gamma(F,G)} \alpha_{n,m}(G(x),F(x))\overset w \rightarrow \sup_{x\in\Gamma(F,G)} 
B_\lambda(G(x),F(x))
\end{equation}
and 
\begin{equation}\label{convupper}
\sup_{x\in\tilde{\Gamma}_{\Delta_{m,n}}(F,G)} \alpha_{n,m}(G(x),F(x))\overset w \rightarrow \sup_{x\in\Gamma(F,G)} 
B_\lambda(G(x),F(x))
\end{equation}
We can assume without loss of generality that $\alpha_{m,1}$ and $\alpha_{n,2}$ are
defined on a rich enough probability space in which there are also independent Brownian bridges,
for which we keep the notation $B_1$, $B_2$ such that $\|\alpha_{m,1}-B_1\|_\infty \to 0$
and $\|\alpha_{n,2}-B_2\|_\infty \to 0$ a.s. (see e.g. Theorem 1, p. 93 in \cite{ShorackWellner}). 
Note that $\|\alpha_{n,m}-B_\lambda\|_\infty \to 0$ a.s., which implies that a.s.
\begin{equation}\label{unifbound}
\sup_{\delta\geq 0} \left|\sup_{x\in\tilde{\Gamma}_\delta(F,G)} \alpha_{n,m}(G(x),F(x)) -\sup_{x\in\tilde{\Gamma}_\delta(F,G)} 
B_\lambda(G(x),F(x)) \right|\leq \|\alpha_{n,m}-B_\lambda\|_\infty \to 0 
\end{equation}
This (take $\delta=0$), proves (\ref{convlower}). We claim that
\begin{equation}\label{aproximacion}
\sup_{x\in\tilde{\Gamma}_{\Delta_{n,m}}(F,G)} B_\lambda(G(x),F(x)) \to \sup_{x\in\Gamma(F,G)}B_\lambda(G(x),F(x)) \mbox{ a.s.}
\end{equation}

In fact, by continuity, $\sup_{x\in\tilde{\Gamma}_{\Delta_{m,n}}(F,G)}B_\lambda(G(x),F(x))=B_\lambda(G(x_n),F(x_n))$ for
some $x_n\in \tilde{\Gamma}_{\Delta_{m,n}}(F,G)$ and by compactness, from
any subsequence we can extract a further subsequence (that we keep
denoting $x_n$) such that $x_n\to x_0$. Since $\Delta_{m,n}\to 0$ a.s., 
necessarily, $x_0\in\Gamma(F,G)$ and
$B_\lambda(G(x_n),F(x_n))\to B_\lambda(G(x_0),F(x_0))$, which means that a.s.
$$\limsup_{n\to\infty}
\sup_{x\in\tilde{\Gamma}_{\Delta_{m,n}}(F,G)} B_\lambda(G(x),F(x))\leq 
\sup_{x\in\Gamma(F,G)}B_\lambda(G(x),F(x)).$$ 
Since, obviously, 
$\sup_{x\in\tilde{\Gamma}_{\Delta{m,n}}(F,G)} B_\lambda(G(x),F(x)))\geq 
\sup_{x\in\Gamma(F,G)}B_\lambda(G(x),F(x))$, we get (\ref{aproximacion}).
Using now (\ref{unifbound}) 
we conclude (\ref{convupper}) and prove (\ref{asympt}).

\bigskip
For the bootstrap result  we note that 
\begin{eqnarray*}
{\textstyle \sqrt \frac {mn}{m+n}} \sup_{x\in\Gamma_{n,m}}((G_m^*(x)-G_m(x))-(F_n^*(x)-F_n(x))))
\overset d =  \sup_{x\in\Gamma_{n,m}} \alpha_{n,m}'(G_m(x),F_n(x)),
\end{eqnarray*}
where $\alpha_{n,m}'$ is an independent copy of $\alpha_{n,m}$ (hence, independent of the $X_i$'s and $Y_j$'s).
We can argue as above and assume that there is an independent copy $B_\lambda$, that we denote $B_\lambda'$ such that
$\|\alpha_{n,m}-B_\lambda'\|_\infty\to 0$ a.s. Since a.s. $B_\lambda'(G_m(x),F_n(x))\to 
B_\lambda'(G(x),F(x))$ we see that we simply have to prove that
\begin{eqnarray*}
V_{n,m}^*&=&\sup_{x\in\Gamma_{n,m}} B_\lambda'(F(x),G(x))\to \sup_{x\in\Gamma(F,G)} B_\lambda'(F(x),G(x)):=V \quad \mbox{ a.s.}
\end{eqnarray*}
To check this, we note that, a.s., $G_m(x)-F_n(x)\to G(x)-F(x)$ uniformly in $x\in\bar{\mathbb{R}}$. 
Consider a sequence of points $x_n\in\Gamma_{n,m}$, that is, such that $G_m(x_n)-F_n(x_n)\geq \pi(F_n,G_m)-\delta_{n,m}$. 
From any subsequence we
can extract a further convergent subsequent for which, again, we keep the notation  
$x_{n}\to x_0$. Then, since $\delta_{n,m}\to0$, $G(x_0)-F(x_0)=
\lim_{n} (G_{m}(x_{n})-F_{n}(x_{n}))\geq \lim_{n} \pi(F_{n},G_{m})-\delta_{n,m}=
\pi(F,G)$, that is, $x_0\in \Gamma(F,G)$. This shows that
$\limsup_{n\to\infty}V_{n,m}^*\leq V$.
For the lower bound, we recall from Lemma \ref{upperlowerbound} that 
for $x\in\Gamma(F,G)$, $G_m(x)-F_n(x)\geq \pi(F_n,G_m)- \Delta_{m,n}$
Now, the choice of $\delta_n$ and the law of iterated logarithm
for the empirical process
(see, e.g., Theorem 1, p. 504 in \cite{ShorackWellner}) ensure 
that a.s.
\begin{eqnarray*}
\limsup_{n\to\infty} \frac {\Delta_{n,m}}{\delta_n}&=&\limsup_{n\to\infty} {\textstyle \frac {2}K}
(\sqrt{\lambda_{n,m}}{\textstyle \frac{\|\alpha_{m,1}\|_\infty}{\sqrt{\log\log m}}}+
\sqrt{1-\lambda_{n,m}}{\textstyle \frac{\|\alpha_{n,2}\|_\infty}{\sqrt{\log\log n}}})\\
&=&{\textstyle \frac{\sqrt{2}}{K}}(\sqrt{\lambda}+\sqrt{1-\lambda})\leq{\textstyle \frac{{2}}{K}}< 1.
\end{eqnarray*}
Hence, eventually $\Delta_{n,m}<\delta_{n,m}$ and if
$x\in\Gamma(F,G)$ then $G_m(x)-F_n(x)\geq \pi(F_n,G_m)-\delta_n$, that is, eventually
$\Gamma(F,G)\subset\Gamma_{n,m}$. As a consequence we see that, with probability one,
$\liminf_{n\to\infty} V_{n,m}^*\geq V$.
This completes the proof. 
\quad $\Box$

\bigskip
Next, we prove the results connected to the limiting distribution in Theorem \ref{Ragha}.

\medskip
\noindent
\textbf{Proof of Proposition \ref{quantilebounds}.} The upper bound for $K_\alpha(F,G,\lambda)$ follows from the obvious fact
\begin{eqnarray*}
\bar{B}(F,G,\lambda)&=&\sup_{t \in T(F,G,\pi(F,G))}\left(\sqrt{ \lambda} \ 
B_1(t)- \sqrt{1-\lambda} \ B_2(t-\pi(F,G))\right)\\
&\leq &\sup_{t \in [\pi(F,G),1]}\left(\sqrt{ \lambda} \ 
B_1(t)- \sqrt{1-\lambda} \ B_2(t-\pi(F,G))\right)=\bar{B}(\pi(F,G),\lambda).
\end{eqnarray*}
For the lower bound note that for every $t\in T(F,G,\pi(F,G))$ we have
$\bar{B}(F,G,\lambda)\geq \sqrt{ \lambda} \ 
B_1(t)- \sqrt{1-\lambda} \ B_2(t-\pi(F,G))$ and this last variable is centered,
normally distributed with variance
$\sigma^{2}_t$
and its $\alpha$-quantile is, therefore, $\sigma_t\Phi^{-1}(\alpha)$. If $\alpha\geq \frac 1 2$
then the best lower bound of this kind is obtained for $\sigma_t=\bar{\sigma}(F,G,\pi(F,G))$,
while for $\alpha< \frac 1 2$ we have $\Phi^{-1}(\alpha)<0$ and the largest upper bound
is given by $\underline{\sigma}(F,G,\pi(F,G)) \Phi^{-1}(\alpha)$. \hfill $\Box$

\bigskip

\medskip
\noindent
\textbf{Proof of Proposition \ref{DistRepre}.} 
We observe first that
$\{\sqrt{ \lambda} \ 
B_1(t)- \sqrt{1-\lambda} \ B_2(t-a)\}_{a\leq t\leq 1}$ has the same distribution as 
$$\textstyle \{\sqrt{1-a} B(\frac{t-a}{1-a})+\sqrt{\lambda a (1-a)}X (1-\frac{t-a}{1-a})+
\sqrt{(1-\lambda)a(1-a)} Y \frac{t-a}{1-a}) \}_{a\leq t\leq 1},$$ where
$B$ is another Brownian bridge and $X$ and $Y$ are independent standard normal r.v.'s, independent
of $B$ (just note that both processes are centered Gaussian with the same
covariance function, namely, $\lambda s(1-t)+(1-\lambda) (s-a)(1-t+a)$ for $a\leq s\leq t\leq 1$).
This implies that
\begin{equation}\label{distequal}
\bar{B}(a,\lambda)\overset d= \sqrt{1-a}\sup_{0\leq s\leq 1}\left(B(s) +\sqrt{\lambda a}(1-s)X+\sqrt{(1-\lambda) a }s Y  \right)
\end{equation}
with $B,X$ and $Y$ as above. 
From this point, we focus, for simplicity, on the case $\lambda=\frac 1 2$, the general 
case following with straighforward but tedious, changes from this.
Using the well-known fact that
\begin{equation}\label{HajekSidak}
P\left(\sup_{0\leq t\leq 1} \left(B(t)-\alpha(1-t)-\beta t\right)>0\right)=
\left\{
\begin{matrix}
e^{-2\alpha \beta} & & \mbox{ if } \alpha>0 , \beta>0\\
1 & & \mbox{ otherwise }
\end{matrix}
\right. ,
\end{equation}
see, e.g., \cite{HajekSidak}, p. 219, we see that
\begin{eqnarray*}
\lefteqn{P(\bar{B}(a,\lambda)>\sqrt{1-a}u)= P \left(\sup_{0\leq s\leq 1}\left(B(s) +\sqrt{a/2}(1-s)X+\sqrt{a/2 }s Y  \right)>u \right)}
\\
&=&1-\Phi({\textstyle \frac{\sqrt{2} u}{\sqrt{a}}})^2 +\int_{x\leq \frac {\sqrt{2}u} {\sqrt{a}},y\leq \frac {\sqrt{2}u} {\sqrt{a}}} 
e^{-2(u-\sqrt{\frac{a}2}x)(u-\sqrt{\frac  a 2}y)} 
{\textstyle \frac 1 {2\pi}} e^{-\frac{x^2+y^2}2}dxdy\\
&=&1-\Phi({\textstyle \frac{\sqrt{2} u}{\sqrt{a}}})^2+
e^{-\frac{2u^2}{1+a}}\int_{-\infty}^{\frac{\sqrt{2}u}{\sqrt{a}}}  {\textstyle \frac 1 {\sqrt{2\pi}}} e^{-\frac{1-a^2}{2}(x-\frac{\sqrt{2a}u}
{1+a})^2} \Phi\left(
{\textstyle \frac{\sqrt{2}u (1-a)}{\sqrt{a}}+ax}
\right)dx
\end{eqnarray*}
and conclude (a). To prove (b), we write 
$$U=\sup_{0\leq t\leq 1} (B(t)+\alpha(1-t)+\beta t).$$
and note from (\ref{HajekSidak}) that $P(U>u)=e^{-2(u-\alpha)(u-\beta)}$ for $u\geq \max(\alpha,\beta)$ and
$P(U>u)=1$ otherwise. Hence, $U$ has density $2(2u-(\alpha+\beta))e^{-2(u-\alpha)(u-\beta)}$,
$u\geq \max(\alpha,\beta)$ and if we write $M(t)=E(e^{tU})$ for the moment generating function of $U$,
then with the change of variable $u=v+\frac{\alpha+\beta}2$ we obtain
\begin{eqnarray*}
M(t)&=&\int_{\alpha\vee \beta}^\infty 2 (2u-(\alpha+\beta)) e^{-2(u-\alpha)(u-\beta)}e^{tu}du\\
&=&e^{\frac{(\alpha-\beta)^2+t(\alpha+\beta)+\frac{t^2}4 }2} \int_{\frac {|\alpha-\beta|}2}^\infty 4v e^{-2(v-\frac{t}4)^2}dv\\
&=&e^{(\alpha\vee \beta)t} e^{\frac 1 2 (|\alpha-\beta|-\frac t 2)^2}\Big[\int_{\frac {|\alpha-\beta|}2}^\infty 4(v-{\textstyle \frac t 4}) 
e^{-2(v-\frac{t}4)^2}dv+t \int_{\frac {|\alpha-\beta|}2}^\infty 
e^{-2(v-\frac{t}4)^2}dv\Big]\\
&=&e^{(\alpha\vee \beta)t}\left[ 1+t {\textstyle \sqrt{\frac{\pi}2}}e^{\frac 1 2 (|\alpha-\beta|-\frac t 2)^2} (1-\Phi(|\alpha-\beta|-
{\textstyle \frac t 2}))\right]\\
&=&e^{(\alpha\vee \beta)t}\left[ 1+{ \frac t 2} 
\frac{(1-\Phi(|\alpha-\beta|-
{\textstyle \frac t 2}))}{\varphi(|\alpha-\beta|-
{\textstyle \frac t 2})}\right].
\end{eqnarray*}
Differentation in this last expression yields now
$$\textstyle E(U)=M'(0)=\alpha\vee \beta+\frac 1 2 \frac{(1-\Phi(|\alpha-\beta|))}{\varphi(|\alpha-\beta|)},$$
$$\textstyle E(U^2)=M''(0)=(\alpha\vee \beta)^2+\frac 1 2+(\alpha+\beta)
\frac 1 2 \frac{(1-\Phi(|\alpha-\beta|))}{\varphi(|\alpha-\beta|)},$$
Now, taking $\alpha=\sqrt{a/2}X$, $\beta=\sqrt{a/2}Y$ and
taking expectacions in the resulting expression we obtain 
\begin{equation}\label{penultima}
E(\bar{B}(a,{\textstyle \frac{1}2}))={\textstyle \sqrt{\frac {a(1-a)}2}} E(\max(X,Y))+{\textstyle \sqrt{\frac {\pi (1-a)}2}}
E(e^{\frac{aZ^2}2} (1-\Phi(\sqrt{a}|Z|))),
\end{equation}
where $Z=(X-Y)/\sqrt{2}$ is standard normal. Observe now that
\begin{eqnarray*}
\lefteqn{E(e^{\frac{aZ^2}2} (1-\Phi(\sqrt{a}|Z|)))=
{\textstyle \frac{2}{\sqrt{2\pi}} }\int_0^\infty e^{-\frac{(1-a)z^2}2 } (1-\Phi(\sqrt{a}z))dz}
\hspace*{1cm}\\
&=&{\textstyle\frac{2}{2\pi} }\int_{(0<\sqrt{a}z<y)} e^{-\frac{(1-a)z^2+y^2}2 }  dzdy
={\textstyle\frac{1}{\pi\sqrt{1-a}} } \int_{(0<\sqrt{a/(1-a)}x<y)} e^{-\frac{x^2+y^2}2 }  dxdy\\
&=&{\textstyle\frac{1}{\pi\sqrt{1-a}} } \int_{(0<r<\infty,\mbox{\scriptsize atan} 
\left({\scriptstyle \sqrt{ \frac {a}{1-a}} }\right)<\theta<\frac \pi 2 )} r e^{-\frac{r^2}2} drd\theta
={\textstyle\frac{1}{\pi\sqrt{1-a}} }\left[{\textstyle\frac{\pi}2} -\mbox{ atan} 
\left({\textstyle \sqrt{ \frac {a}{1-a}} }\right) \right].
\end{eqnarray*}
Plugging this into (\ref{penultima}) and taking into account that $E(\max(X,Y))=\frac 1 {\sqrt{\pi}}$ 
we obtain the conclusion about $E(\bar{B}(a,{\textstyle \frac{1}2}))$. 
A similar computation yields $E(\bar{B}(a,{\textstyle \frac{1}2}))=\frac{1-a^2}2$ and completes the
proof. \hfill $\Box$

\medskip
Next, we prove the result about the level and power of the test
for essential stochastic order.

\medskip
\noindent
\textbf{Proof of Proposition \ref{Test1}.} We assume for simplicity $m=n$. The general case can be handled with straighforward
changes. A simple computation shows that $\pi(F_0,G_0)=\pi_0$ and $\Gamma(F_0,G_0)=\{\frac {1+\pi_0}2 \}$
and, using Theorem \ref{Ragha}, that $\sqrt{\frac n 2} (\pi_{n,n}-\pi_0)\overset w \to N(0,\bar{\sigma}_{\pi_0}^2)$.
Hence, 
\begin{eqnarray*}
\lefteqn{\liminf_{n\to\infty} \sup_{(F,G)\in H_0} \mathbb{P}_{F,G}(  \textstyle{\sqrt{\frac{n}{2}}
(\pi_{n,n}-\pi_0) < \bar{\sigma}_{\pi_0}\Phi^{-1}(\alpha)} )}\hspace*{1cm}\\
&\geq& \lim_{n\to\infty}  \mathbb{P}_{F_0,G_0}(  \textstyle{\sqrt{\frac{n}{2}}
(\pi_{n,n}-\pi_0) < \bar{\sigma}_{\pi_0}\Phi^{-1}(\alpha)} ) =\alpha.
\end{eqnarray*}
For the upper bound we recall from (\ref{mainupperlowerbound}) that 
$\pi_{n,n}-\pi(F,G)\geq  \alpha_{n,n}(G(x),$ $G(x)-\pi(F,G))$
for every $x\in\Gamma(F,G)$. As a consequence, for $(F,G)\in H_0$ and $x\in\Gamma(F,G)$,
\begin{eqnarray}
\nonumber
\lefteqn{\mathbb{P}_{F,G}(  \textstyle{\sqrt{\frac{n}{2}}
(\pi_{n,n}-\pi_0) \leq \bar{\sigma}_{\pi_0}\Phi^{-1}(\alpha)} )} \hspace*{0.5cm}\\
\label{preHoeff}
&\leq &
P\left(\alpha_{n,n}(G(x),G(x)-\pi(F,G)) \leq \bar{\sigma}_{\pi_0}\Phi^{-1}(\alpha)- 
\textstyle{\sqrt{\frac{n}{2}}(\pi(F,G)-\pi_0)}
\right).
\end{eqnarray}
We observe that, for any $x\in\Gamma(F,G)$, $\alpha_{n,n}(G(x),$ $G(x)-\pi(F,G))$
is a sum of $n$ i.i.d. centered random variables
with variance $\sigma^2(x)=\frac 1 2 G(x)(1-G(x))+\frac 1 2 (G(x)-\pi(F,G))(1-(G(x)-\pi(F,G)))$
and third absolute moment smaller than $2^{3/2}$. From the Berry-Esseen inequality (see, e.g., 
Theorem 1, p. 848 in \cite{ShorackWellner})
we see that for some universal constant $C>0$
\begin{eqnarray*}
\mathbb{P}_{F,G}(  \textstyle{\sqrt{\frac{n}{2}}
(\pi_{n,n}-\pi_0) \leq \bar{\sigma}_{\pi_0}\Phi^{-1}(\alpha)} )\leq \Phi\left({\textstyle \frac{\bar{\sigma}_{\pi_0}\Phi^{-1}(\alpha)
-\sqrt{\frac n 2 }(\pi(F,G)-\pi_0)}{\sigma(x)}}\right)+\frac{2^{3/2}C}{\sigma(x)^{3/2} \sqrt{n}}.
\end{eqnarray*}
The computations leading to the expressions for $\bar{\sigma}^2_\pi$ and 
$\underline{\sigma}_\pi^2$ show that $\frac 1 2 \pi(F,G)(1-\pi(F,G))\leq \sigma(x)\leq \frac 1 4(1-\pi^2(F,G))\leq
\frac 1 4(1-\pi_0)=\bar{\sigma}_{\pi_0}^2$ for $(F,G)\in H_0$ and $x\in\Gamma(F,G)$. This and the fact that 
$\bar{\sigma}_{\pi_0}\Phi^{-1}(\alpha)
-\sqrt{\frac n 2 }(\pi(F,G)-\pi_0)\leq 0$ yield that for every $(F,G)\in H_0$
\begin{eqnarray*}
\mathbb{P}_{F,G}(  \textstyle{\sqrt{\frac{n}{2}}
(\pi_{n,n}-\pi_0) \leq \bar{\sigma}_{\pi_0}\Phi^{-1}(\alpha)} )& \leq &\Phi\left( {\textstyle \Phi^{-1}(\alpha)
-\frac {\sqrt{n}} {2\bar{\sigma}_{\pi_0}}(\pi(F,G)-\pi_0) }\right) \\ 
&&+ \frac{8C}{(\pi(F,G)(1-\pi(F,G)))^{3/2} \sqrt{n}}\\
&\leq& \alpha +\frac{8C}{(\pi(F,G)(1-\pi(F,G)))^{3/2} \sqrt{n}}.
\end{eqnarray*}
On the other hand, from (\ref{preHoeff}) and Hoeffding's inequality we see that 
\begin{eqnarray}\nonumber
\mathbb{P}_{F,G}(  \textstyle{\sqrt{\frac{n}{2}}
(\pi_{n,n}-\pi_0) \leq \bar{\sigma}_{\pi_0}\Phi^{-1}(\alpha)} )&\leq & e^{-2(
\bar{\sigma}_{\pi_0}\Phi^{-1}(\alpha)
-\sqrt{\frac n 2 }(\pi(F,G)-\pi_0))^2}\\
\label{perrortipoI}
&\leq &
e^{-n(\pi(F,G)-\pi_0)^2}.
\end{eqnarray}
This, in particular, yields (\ref{errortipo1}). 
Now fix $\delta>0$ small enough to ensure that $\pi_0+\delta<1$ and 
$\pi(1-\pi)\geq \frac 1 {2^{2/3}} \pi_0(1-\pi_0)$
if $\pi_0+\delta\geq\pi\geq \pi_0$. Then
\begin{eqnarray*}
\sup_{(F,G)\in H_0} \mathbb{P}_{F,G}(  \textstyle{\sqrt{\frac{n}{2}}
(\pi_{n,n}-\pi_0) \leq \bar{\sigma}_{\pi_0}\Phi^{-1}(\alpha)} )\leq 
\alpha + \frac{16 C}{ (\pi_0(1-\pi_0))^{3/2} \sqrt{n}} + e^{-\frac{n}2 \delta}\to \alpha
\end{eqnarray*}
and this proves (\ref{asymptoticlevel}).

\medskip
Finally, for the proof of (\ref{power}) we simply note that
$\pi_{n,n}-\pi(F,G)\leq n^{-1/2}(\|\alpha_{n,1}\|_\infty+\|\alpha_{n,2}\|_\infty)$
and, therefore,
\begin{eqnarray*}
\lefteqn{\mathbb{P}_{F,G}(  \textstyle{\sqrt{\frac{n}{2}}
(\pi_{n,n}-\pi_0) > \bar{\sigma}_{\pi_0}\Phi^{-1}(\alpha)})}\hspace*{0.5cm}\\
&=&
\mathbb{P}_{F,G}(  \textstyle{\sqrt{\frac{n}{2}}
(\pi_{n,n}-\pi(F,G)) > \bar{\sigma}_{\pi_0}\Phi^{-1}(\alpha)}+\textstyle{\sqrt{\frac{n}{2}}
(\pi_0-\pi(F,G))})\\
&\leq &P(\|\alpha_{n,1}\|_\infty >K/\sqrt{2})+
P(\|\alpha_{n,2}\|_\infty >K/\sqrt{2}),
\end{eqnarray*}
where $K=\bar{\sigma}_{\pi_0}\Phi^{-1}(\alpha)+\sqrt{\frac n 2}(\pi_0-\pi(F,G))$. An application of the Dvoretzky-Kiefer-Wolfowitz
inequality, see \cite{Massart}, yields
$$\mathbb{P}_{F,G}(  \textstyle{\sqrt{\frac{n}{2}}
(\pi_{n,n}-\pi_0) > \bar{\sigma}_{\pi_0}\Phi^{-1}(\alpha)})\leq 2 e^{-K^2}$$
and completes the proof.\hfill $\Box$

\medskip
\noindent
\textbf{Proof of Proposition \ref{vanishingbias}.} 
We keep the notation of the proof of Theorem \ref{Ragha} with
$G_m(x)=G(x)+\frac{1}{\sqrt{m}}\alpha_{m,1}(G(x))$
and $F_n(x)=F(x)+\frac{1}{\sqrt{n}}\alpha_{n,2}(F(x))$ for independent uniform empirical processes $\alpha_{m,1}$,
$\alpha_{n,2}$ that we assume, without loss of generality, to be defined on a rich enough probability
space in which there are independent Brownian bridges, $B_{m,1}$,
$B_{n,2}$, satisfying
\begin{equation}\label{KMT}
P(\|\alpha_{m,1}-B_{m,1}\|_\infty>n^{-1/2}(x+12\log n))\leq 2 e^{-x/6},\quad x>0
\end{equation}
and similarly for $\alpha_{n,2}$ and $B_{n,2}$ (see, e.g., \cite{CsorgoHorvath}, p. 114). 
In particular, we have that $E(\|\alpha_{m,1}-B_{m,1}\|_\infty)\leq \frac{12(1+\log m)}{\sqrt{m}}$.
We define $\tilde{G}_m(x)=G(x)+\frac{1}{\sqrt{m}}B_{m,1}(G(x))$,
$\tilde{F}_n(x)=F(x)+\frac{1}{\sqrt{n}}B_{n,2}(F(x))$ and $\pi(\tilde{F}_n,\tilde{G}_m)=
\sup_{x\in\mathbb{R}} (\tilde{G}_m(x)-\tilde{F}_n(x))$. From (\ref{KMT}) we obtain that
\begin{eqnarray}\nonumber
\lefteqn{\textstyle \sqrt{\frac{nm}{n+m}}E| \pi(F_n,G_m)-\pi(\tilde{F}_n,\tilde{G}_m)|}\hspace*{1cm}\\
\label{espacero}
&\leq &\sqrt{\lambda_{n,m}}E(\|\alpha_{m,1}-B_{m,1}\|_\infty)+
\sqrt{1-\lambda_{m,n}} E(\|\alpha_{n,2}-B_{n,2}\|_\infty)\to 0,
\end{eqnarray}
as $n,m\to\infty$.
We can write, as well, $G_m^*(x)=G_m(x)+\frac 1 {\sqrt{m}}\alpha_{m,2}'(G_m(x))$,  $F_n^*(x)=F_n(x)+\frac 1 {\sqrt{n}}
\alpha_{n,2}'(F_n(x))$ with $\alpha_{m,1}'$, $\alpha_{n,2}'$ independent uniform empirical
processes, independent of $\alpha_{m,1}$ and $\alpha_{n,2}$ and $B_{m,1}'$, $B_{n,2}'$ for Brownian bridges
related to $\alpha_{m,1}'$, $\alpha_{n,2}'$ as in (\ref{KMT}). Now, we define
$\tilde{G}_m^*(x)=\tilde{G}_m(x)+\frac 1 {\sqrt{m}}\alpha_{m,1}'(G_m(x))$,
$\tilde{F}_n^*(x)=\tilde{F}_n(x)+\frac 1 {\sqrt{n}}
\alpha_{n,2}'(\tilde{F}_n(x))$, $\pi(\tilde{F}_n^*,\tilde{G}_m^*)=
\sup_{x\in\mathbb{R}} (\tilde{G}_m^*(x)-\tilde{F}_n^*(x))$.
We claim that 
\begin{equation}\label{espbcero}
\textstyle \sqrt{\frac{nm}{n+m}}E| \pi(F_n^*,G_m^*)-\pi(\tilde{F}_n^*,\tilde{G}_m^*)|\to 0.
\end{equation}
In fact, after estimate (\ref{espacero}) it suffices to show that
\begin{equation}\label{espccero}
E\left(\sup_{x} |\alpha_{m,1}'(G_m(x))- \alpha_{m,1}'(\tilde{G}_m(x))| \right)\to 0
\end{equation}
and the same for $\alpha_{n,2}'$.
But we have  that $E\sup_{0\leq t-s \leq a} |B'_{m,1}(t)-B'_{m,1}(s)|\leq 7 \sqrt{2 a \log(1/a)}$, for $a\in (0,\frac 1 2]$
(see Theorem 3, p. 538, in \cite{ShorackWellner}). Hence, conditioning on $G_m,\tilde{G}_m$ and taking $\nu\in (0,\frac 1 2)$  
we obtain from (\ref{KMT}) that from some positive constant, $K$,
$$E\left(\sup_{x} |\alpha_{m,1}'(G_m(x))- \alpha_{m,1}'(\tilde{G}_m(x))|\right)\leq K \left(\textstyle \frac{\log m}{\sqrt{m}}
+\frac 1 {m^{\nu/2}}E(\|\alpha_{m,1}-B_{m,1}\|^{s}_\infty)\right)\to 0.$$
Now, from (\ref{espacero}) and (\ref{espbcero}) we see that the result will follow if we prove
that
\begin{equation}\label{versionequivalente}
\textstyle \sqrt{\frac{nm}{n+m}}E| \pi(\tilde{F}^*_n,\tilde{G}^*_m)-\pi(\tilde{F}_n,\tilde{G}_m)|\to 0.
\end{equation}
As in the proof of Theorem \ref{Ragha} we see that
$$\textstyle \sup_{x\in\Gamma(\tilde{F}_n,\tilde{G}_m)} \alpha_{n,m}'(x)\leq{\textstyle 
\sqrt{\frac{nm}{n+m}}(\pi(\tilde{F}^*_n,\tilde{G}^*_m)-\pi(\tilde{F}_n,\tilde{G}_m))}\leq
\sup_{x\in \bar{\Gamma}(\tilde{F}_n,\tilde{G}_m)} \alpha_{n,m}'(x),
$$
where $\alpha_{n,m}'(x)=\sqrt{\lambda_{n,m}} \alpha_{m,1}'(\tilde{G}_m(x))+
\sqrt{1-\lambda_{n,m}} \alpha_{n,2}'(\tilde{F}_n(x))$, $\Gamma(\tilde{F}_n,\tilde{G}_m)=\{x: \tilde{G}_m(x)$ $-\tilde{F}_n(x)
=\pi(\tilde{F}_n,\tilde{G}_m)\}$ and 
$\bar{\Gamma}(\tilde{F}_n,\tilde{G}_m)=\{x: \tilde{G}_m(x)-\tilde{F}_n(x)
\geq \pi(\tilde{F}_n,\tilde{G}_m)-2(\|\alpha'_{m,1}\|_\infty/$ $\sqrt{m}+\|\alpha'_{n,2}\|_\infty/\sqrt{n})$.
We can mimmick the argument in Theorem \ref{Ragha} to show that
$$\sup_{x\in\bar{\Gamma}(\tilde{F}_n,\tilde{G}_m)} \alpha_{n,m}'(x)-\sup_{x\in\Gamma(\tilde{F}_n,\tilde{G}_m)} \alpha_{n,m}'(x)\to 0$$
in $L_1$. Finally, to show that
$$E\textstyle \Big({\displaystyle \sup_{x\in\Gamma(\tilde{F}_n,\tilde{G}_m)}} \alpha_{n,m}'(x)\Big)\to 0,$$
observe that $\tilde{G}_m(x)-\tilde{F}_n(x)$ is a Gaussian process with continuous sample paths whose increments
have nonzero variance. As a consequence (see
Lemma 2.6 in \cite{KimPollard}), with probability one, $\Gamma(\tilde{F}_n,\tilde{G}_m)$ consists of just one point, say
$x_{n,m}$, which depends on $\tilde{F}_n$ and $\tilde{G}_m$. Conditionally given 
$\tilde{F}_n$ and $\tilde{G}_m$, $\sup_{x\in\Gamma(\tilde{F}_n,\tilde{G}_m)} \alpha_{n,m}'(x)=\alpha_{n,m}'(x_{n,m})$
is a centered random variable. But taking expectations we see that, in fact,
$$E\textstyle \Big({\displaystyle \sup_{x\in\Gamma(\tilde{F}_n,\tilde{G}_m)}} \alpha_{n,m}'(x)\Big)=0.$$
This completes the proof. \hfill $\Box$

\medskip
\noindent \textbf{Proof of Proposition \ref{TestAgainst}.} We only deal with (\ref{asymptoticlevelag})
since (\ref{errortipo1ag}) follows from the Dvoretzky-Kiefer-Wolfowitz inequality
and (\ref{errortipo2ag}) from Hoeffding's inequality reproducing almost verbatim
the arguments in Proposition \ref{Test1}. Also, for ease of notation, we consider the
case $m=n$. For the pair $F_0,G_0$ we have $\pi(F_0,G_0)=\pi_0$ and $T(F_0,G_0,\pi_0)=[\pi_0,1]$.
This and Theorem \ref{Ragha} imply that
\begin{eqnarray*}
\lefteqn{\liminf_{n\to\infty} \sup_{(F,G)\in H_0}\mathbb{P}_{F,G}({\textstyle{\sqrt{\frac n 2}}} 
(\pi_{n,n}-\pi_0)>K_{1-\alpha}(\pi_0,{\textstyle \frac 1 2}))}\hspace*{3cm}
\\
&\geq& 
\lim_{n\to\infty}
\mathbb{P}_{F_0,G_0}({\textstyle{\sqrt{\frac n 2}}} 
(\pi_{n,n}-\pi_0)>K_{1-\alpha}(\pi_0,{\textstyle \frac 1 2}))=\alpha.
\end{eqnarray*}
To complete the proof of (\ref{asymptoticlevelag}) assume, without loss of generality, 
that, as in the proof of Proposition \ref{vanishingbias}, $\alpha_{n,1}$ and $\alpha_{n,2}$ are defined
on a rich enough probability space together with Brownian bridges $B_{n,1},B_{n,2}$ 
satisfying (\ref{KMT}). In particular, if $B_n(s,t)=\frac 1 2 B_{n,1}(s)+\frac 1 2 B_{n,2}(t)$, then
there are universal constants $c_1,c_2>0$ such that 
\begin{equation}\label{KMT2}
P(\|\alpha_{n,n}-B_n\|\geq c_1{\textstyle \frac{\log n}{\sqrt{n}}})\leq {\textstyle \frac {c_2}{n^2}} .
\end{equation}
Recall from Lemma \ref{upperlowerbound} 
that 
\begin{equation}\label{otracota}
{\textstyle{\sqrt{\frac n 2}}} 
(\pi_{n,n}-\pi_0)\leq \sup_{t\in [\pi(F,G),1]} \alpha_{n,n}(t,t-\pi(F,G))+R_{n,n}
\end{equation}
with $R_{n,n}=\frac{1}{\sqrt{2}}(\omega_{n,1}(\Delta_{n,n})+\omega_{n,2}(\Delta_{n,n}))$.
We saw in the proof of Theorem \ref{Ragha} that $\limsup_{n\to\infty} \frac{\Delta_{n,n}}{\delta_n}= \frac 2 K<1$
a.s. if $\delta=K\sqrt{\frac 2{n} \log \log n}$ and $K>2$. This implies that a.s., eventually $\omega_{n,1}(\Delta_{n,n})
\leq\omega_{n,1}(\delta_n)$. From Stute's results on the oscillation of the empirical process (see, e.g. Theorem 1, p.
542 in \cite{ShorackWellner}) we have that a.s.
$$\lim_{n\to\infty}\frac{\sqrt{n}\omega_{n,1}(\delta_n)}{\sqrt{K\sqrt{2} \log n \log\log n}}=1.$$
Consequently, $\frac{\sqrt{n}\omega_{n,1}(\Delta_{n,n})}{\log n}\to 0$ a.s. 
and the same happens for $\omega_{n,2}(\Delta_{n,n})$. Hence, $\frac{\sqrt{n}R_{n,n}}{\log n}\to 0$ a.s. and, in particular
\begin{equation}\label{oscillationbound}
P(R_{n,n}>{\textstyle \frac{\log n}{\sqrt{n}}})\to 0.
\end{equation}
Now, combining (\ref{otracota}), (\ref{KMT2}) and (\ref{oscillationbound}) we obtain that
\begin{eqnarray*}
\lefteqn{\mathbb{P}_{F,G}({\textstyle{\sqrt{\frac n 2}}} 
(\pi_{n,n}-\pi_0)>K_{1-\alpha}(\pi_0,{\textstyle \frac 1 2}))}\hspace*{0.5cm}
\\ &\leq &
P\Big(\sup_{t\in [\pi(F,G),1]} \alpha_{n,n}(t,t-\pi(F,G))+R_{n,n} > K_{1-\alpha}
(\pi_0,{\textstyle \frac 1 2})+{\textstyle{\sqrt{\frac n 2}}} 
(\pi_0-\pi(F,G)) \Big)\\
\\ &\leq &
P\Big(\sup_{t\in [\pi(F,G),1]} B_n(t,t-\pi(F,G))> K_{1-\alpha}
(\pi_0,{\textstyle \frac 1 2})-{\textstyle \frac{(c_1+1)\log n}{\sqrt{n}}}+{\textstyle{\sqrt{\frac n 2}}} 
(\pi_0-\pi(F,G)) \Big)\\
&&+P(R_{n,n}>{\textstyle \frac{\log n}{\sqrt{n}}})+{\textstyle \frac{c_2}{n^2}}.
\end{eqnarray*}
This shows that if suffices to prove that
\begin{equation}\label{finalclaim}
\limsup_{n\to\infty}\sup_{a \leq \pi_0 } P\Big(\bar{B}(a,{\textstyle \frac{1}{2}})> K_{1-\alpha}
(\pi_0,{\textstyle \frac 1 2})+{\textstyle{\sqrt{\frac n 2}}} 
(\pi_0-a)-r_n\Big)\leq \alpha
\end{equation}
if $r_n\searrow 0$. To check this we note that the distribution function of $P(\bar{B}(a,{\textstyle \frac{1}{2}})\leq x)$
depends continuously on $(a,x)$ (this follows easily from Proposition \ref{DistRepre} (a), for instance).
Hence, given $\varepsilon>0$ we can find $\pi_1<\pi_0$ and $\delta>0$ such that 
$P\Big(\bar{B}(a,{\textstyle \frac{1}{2}})> K_{1-\alpha}
(\pi_0,{\textstyle \frac 1 2})-r \Big)\leq \alpha+\varepsilon$ if $\pi_1\leq a\leq \pi_0$ and $0\leq r\leq\delta$.
But then, taking $n$ large enough to ensure that $r_n\leq \delta$ we have
\begin{eqnarray*}
\lefteqn{\sup_{\pi_1\leq a \leq \pi_0 } P\Big(\bar{B}(a,{\textstyle \frac{1}{2}})> K_{1-\alpha}
(\pi_0,{\textstyle \frac 1 2})+{\textstyle{\sqrt{\frac n 2}}} 
(\pi_0-a)-r_n\Big)}\hspace*{3cm}\\
&\leq& 
\sup_{\pi_1\leq a \leq \pi_0 } P\Big(\bar{B}(a,{\textstyle \frac{1}{2}})> K_{1-\alpha}
(\pi_0,{\textstyle \frac 1 2})-\delta\Big)
\leq \alpha+\epsilon,
\end{eqnarray*}
while
\begin{eqnarray*}
\lefteqn{\sup_{a \leq \pi_1 } P\Big(\bar{B}(a,{\textstyle \frac{1}{2}})> K_{1-\alpha}
(\pi_0,{\textstyle \frac 1 2})+{\textstyle{\sqrt{\frac n 2}}} 
(\pi_0-a)-r_n\Big)}\hspace*{3cm}\\
&\leq& 
\sup_{a \leq \pi_1 } P\Big(\bar{B}(a,{\textstyle \frac{1}{2}})> {\textstyle \sqrt{\frac n 2}}(\pi_0-\pi_1) \Big)\\
&\leq & P\Big(\|B_1\|_\infty+\|B_2\|_\infty> {\textstyle \sqrt{n }}(\pi_0-\pi_1) \Big)\\
&\leq & 2 e^{-\frac n 2 (\pi_0-\pi_1)^2}.
\end{eqnarray*}
The last two estimates complete the proof. \hfill $\Box$

 \end{document}